\documentclass[conference]{IEEEtran}
\IEEEoverridecommandlockouts
\usepackage{cite}
\usepackage{amsmath,amssymb,amsfonts}
\usepackage{algorithmic}
\usepackage{booktabs}
\usepackage{multirow}
\usepackage[normalem]{ulem}
\useunder{\uline}{\ul}{}
\usepackage{graphicx}
\usepackage{textcomp}
\usepackage{xcolor}
\usepackage{etoolbox}

\def\BibTeX{{\rm B\kern-.05em{\sc i\kern-.025em b}\kern-.08em
    T\kern-.1667em\lower.7ex\hbox{E}\kern-.125emX}}

\usepackage{enumitem}
\usepackage{makecell}
\usepackage{hyperref}
\usepackage{threeparttable}
\usepackage{subfig}

\usepackage{color, xspace}

\newcommand{\ie}{\emph{i.e.,}\xspace}
\newcommand{\eg}{\emph{e.g.,}\xspace}
\newcommand{\etal}{\emph{et al.}\xspace}
\newcommand{\mname}{Uni-CTR\xspace}

\begin{document}

\title{
A Unified Framework for Multi-Domain CTR Prediction via Large Language Models
}

\author{
    \IEEEauthorblockN{Zichuan Fu$^{*}$\thanks{*Co-first authors with equal contributions.}}
    \IEEEauthorblockA{
        \textit{City University of Hong Kong}\\
        zc.fu@my.cityu.edu.hk \\
    }
    \and
    \IEEEauthorblockN{Xiangyang Li$^{*}$}
    \IEEEauthorblockA{
        \textit{Huawei Noah’s Ark Lab}\\
        lixiangyang34@huawei.com
    }
    \and
    \IEEEauthorblockN{Chuhan Wu$^{*}$}
    \IEEEauthorblockA{
        \textit{Huawei Noah’s Ark Lab}\\
        wuchuhan1@huawei.com
    }
    \and
    \IEEEauthorblockN{Yichao Wang}
    \IEEEauthorblockA{
        \textit{Huawei Noah’s Ark Lab}\\
        wangyichao5@huawei.com
    }
    \and
    \IEEEauthorblockN{Kuicai Dong}
    \IEEEauthorblockA{
        \textit{Huawei Noah’s Ark Lab}\\
        dong.kuicai@huawei.com
    }
    \and
    \IEEEauthorblockN{Xiangyu Zhao}
    \IEEEauthorblockA{
        \textit{City University of Hong Kong}\\
        xianzhao@cityu.edu.hk
    }
    \and
    \IEEEauthorblockN{Mengchen Zhao}
    \IEEEauthorblockA{
        \textit{Huawei Noah’s Ark Lab}\\
        zhaomengchen@huawei.com
    }
    \and
    \IEEEauthorblockN{Huifeng Guo}
    \IEEEauthorblockA{
        \textit{Huawei Noah’s Ark Lab}\\
        huifeng.guo@huawei.com
    }
    \and
    \IEEEauthorblockN{Ruiming Tang}
    \IEEEauthorblockA{
        \textit{Huawei Noah’s Ark Lab}\\
        tangruiming@huawei.com
    }
}

\maketitle

\begin{abstract}
    Multi-Domain Click-Through Rate (MDCTR) prediction is crucial for online recommendation platforms, which involves providing personalized recommendation services to users in different domains.
    However, current MDCTR models are confronted with the following limitations.
    Firstly, due to varying data sparsity in different domains, models can easily be dominated by some specific domains, which leads to significant performance degradation in other domains (i.e., the ``seesaw phenomenon"). 
    Secondly, when new domain emerges, the scalability of existing methods is limited, making it difficult to adapt to the dynamic growth of the domain. 
    Traditional MDCTR models usually use one-hot encoding for semantic information such as product titles, thus losing rich semantic information and leading to insufficient generalization of the model.
    In this paper, we propose a novel solution Uni-CTR to address these challenges. Uni-CTR leverages Large Language Model (LLM) to extract layer-wise semantic representations that capture domain commonalities, mitigating the seesaw phenomenon and enhancing generalization. Besides, it incorporates a pluggable domain-specific network to capture domain characteristics, ensuring scalability to dynamic domain growth.
    Experimental results on public datasets and industrial scenarios show that Uni-CTR significantly outperforms state-of-the-art (SOTA) models. In addition, Uni-CTR shows significant results in zero shot prediction. Code is available at \href{https://github.com/archersama/Uni-CTR}{https://github.com/archersama/Uni-CTR}.
\end{abstract}


\section{Introduction}
\label{sec:introduction}


Click-Through Rate (CTR) prediction aims to estimate the likelihood of a person engaging with an advertisement or item by clicking on it. This estimation is essential for many online recommendation platforms, including e-commerce, advertising, social networking, and multimedia services \cite{richardson2007predicting, yang2022click}.
The recent advent of digitization has led to the expansion of commercial platforms, diversifying the range of services and products across numerous domains~\cite{digitization}. 
They offer services such as online shopping, ride-sharing, food delivery, and virtual event hosting, providing users with more options and greater convenience~\cite{service}.
Therefore, their recommender systems need to serve various scenarios and generate accurate CTR predictions for them.

Since the data from different domains usually have quite different characteristics, simply mixing multi-domain data is often sub-optimal~\cite{mmoe}.
Thus, traditional single-domain CTR systems \cite{qu2016product, wang2017deep, guo2017deepfm, lian2018xdeepfm, song2019autoint, fibinet} typically train separate models based on data from different domains. 
However, this approach often overlooks the shared characteristics inherent in data across different domains. 
For instance, if users exhibit a preference for purchasing sportswear in online shopping, they are also likely to be interested in sports-related content for video streaming.
Ignoring such shared information can lead to sub-optimal model performance and loss of cross-domain knowledge. Furthermore, training a model for each domain would significantly increase the resource
and manpower costs of the system.

To mitigate this issue, Multi-Domain CTR (MDCTR) systems \cite{juan2016field, mmoe, ple, sheng2021one, yang2022adasparse, chang2023pepnet} have been developed to make accurate predictions across various domains. The ultimate goal of MDCTR systems is to provide personalized recommendation services for users in different domains.  However, the current MDCTR systems encounter challenges in achieving this goal due to the following reasons:

\begin{enumerate}[leftmargin=*,itemsep=0pt,topsep=0pt]

    \item \textbf{Sparsity of Domain Data.} 
    In practice, many domains do not accumulate sufficient training data, and the amount of click feedback is even sparser. Therefore, models can easily be dominated by data-rich domains, which leads to significant performance degradation in other domains (i.e., the ``seesaw phenomenon"). As a result, the performance of conventional MDCTR models in many domains is not satisfactory, especially on long-tailed samples.

    \item \textbf{Limited Scalability.} 
    Traditional multi-domain methods often have to either fine-tune the entire existing model or train a completely new model when a new domain is introduced~\cite{luo2024domainaware}. 
    STAR \cite{sheng2021one} uses topology to address this challenge, which performs an element-wise multiplication operation between the outputs of the backbone network and the domain-specific network to obtain the final output.
    However, such design requires the size and structure of domain-specific networks to be identical to those of the backbone network. Therefore, as the number of domains increases, the number of model parameters increases multiplicatively.
    
    \item \textbf{Weak Generalization.} 
    In traditional multi-domain methods, features are typically transformed into discrete IDs as inputs; such a conversion process often results in the loss of semantic information associated with these data (e.g., computer and keyboard). When encountering a cold-start recommendation situation, the MDCTR model is difficult to quickly generalize to new domains (i.e., zero-shot prediction), leading to sub-optimal results.
\end{enumerate}

To address the above challenges, we propose \mname, a unified framework for Multi-Domain CTR Prediction that incorporates large language models as bridges among different domains. 
The foundation of \mname is using natural language as a general information carrier so that knowledge from different domains can be universally encoded and exploited.
Specifically, \mname employs a well-curated \textit{prompting strategy} to convert non-textual and textual features into a correlated prompt sequence, which preserves rich semantic and contextual information.
Secondly, \mname utilizes an \textit{LLM backbone} to capture commonalities of all domains, the incorporated \textit{domain-specific networks} (DSN) to learn characteristics of different domains. Specifically, DSN can utilize contextual knowledge from various layers of LLM since the bottom layers of the LLM learn surface phrase-level features, while higher ones focus more on understanding more complex and semantic concepts \cite{jawahar2019does}.
Besides, the DSN network is pluggable and can be flexibly added or removed when new domains emerge or old domains become obsolete, greatly increasing the scalability of the model. Specifically, we design a \textit{masked loss strategy} by masking the loss back-propagated to different parts of \mname networks. The gradient of each sample is used to update: (i) its corresponding domain-specific network only, rather than all domain-specific networks, and (ii) the LLM backbone and the general network. 
Our masked loss strategy ensures the decoupling between domain-specific networks. It has two advantages: (i) alleviating the domain seesaw problem because common and distinct features are modeled separately, and (ii) improving the system's scalability as each domain-specific network becomes pluggable with respect to the LLM backbone. Therefore, while training a new domain or fine-tuning a specific existing domain, the parameters of other domain-specific networks will not be affected. Lastly, a \textit{general network} to further refine the commonalities for zero-shot prediction.

In summary, our contributions are three-fold:
\begin{itemize}[leftmargin=*]

    \item We propose \mname, an LLM-based model, for multi-domain CTR prediction. It utilizes domain-specific networks to capture domain characteristics and LLM to capture domain commonalities greatly mitigates the seesaw phenomenon.  To the best of our knowledge, this is the first utilization of a large language model as the backbone for MDCTR prediction.

    \item We introduce a masked loss strategy to ensure the decoupling of domain-specific networks relative to the LLM backbone. This allows DSNs to be pluggable so they can be flexibly added or removed as new domains are added or old domains are removed.
    
    \item Experimental results indicate that \mname outperforms the baseline with an impressive margin regarding prediction accuracy and zero-shot prediction capabilities. This holds for both public and industrial datasets. These findings confirm the effectiveness of our approach and its potential for wider application in MDCTR prediction.
    
\end{itemize}

The following of this paper is organized as follows: in Section~\ref{sec:related_work}, we provide an overview of the existing research on traditional single-domain and multi-domain CTR prediction, and LLM-based CTR models. A preliminary background of MDCTR prediction is presented in Section~\ref{sec:task-formulation}. Section~\ref{sec:method} elaborates on the architecture of \mname. Section~\ref{sec:experiments} delves into the experimental design and a comparative analysis with existing models, underscoring our model's performance and zero-shot prediction ability. Furthermore, our model is also verified on an industrial dataset, shown in section~\ref{sec:industry}. Finally, Section~\ref{sec:conclusion} wraps up the paper by summarizing our contributions and outlining avenues for future research.

\section{Related Work}
\label{sec:related_work}

\subsection{Click-Through Rate (CTR) prediction}

        Click-through rate (CTR) prediction is a task to forecast the probability that a user will click on a given item, such as an advertisement or a product. 
        Typically, CTR prediction models employ advanced algorithmic models~\cite{svmctr,lrctr,deepctr,convctr,juan2016field,din,guo2017deepfm,attentionctr} to analyze user-item data and contextual information, thereby accurately estimating the likelihood of user engagement with specific content. 
        User and item features for CTR prediction modeling typically include user personal information, browsing and purchasing history, user interaction patterns, and item attributes. These features allow for a subtle and nuanced understanding of user preferences and behaviors, improving the accuracy of personalized content recommendations.
        The evolution of CTR prediction models has progressed from multivariate statistical approaches \cite{regression} to factorization machines (FMs) \cite{rendle2010factorization}, deep learning \cite{emmert2020introductory, deng2018ad} and hybrid techniques, reflecting advancements in capturing complex feature interactions and patterns in increasingly large industry scenario.

        Factorization Machines (FMs) \cite{rendle2010factorization} initially dominated the landscape of CTR prediction models. Their ability to capture interactions between features and computational efficiency made them particularly effective in handling sparse datasets. With the birth of deep neural networks (DNNs), there was a significant shift in CTR prediction methodologies. This era witnessed the emergence of models that synergized FMs and DNNs, harnessing the strengths of both approaches. 
        For example, DeepFM \cite{guo2017deepfm} integrates factorization machines for low-order feature interactions with deep neural networks for high-order feature interactions, offering computational efficiency and superior performance across various tasks. xDeepFM \cite{lian2018xdeepfm} further innovates by introducing the Compressed Interaction Network (CIN) to capture high-order feature interactions efficiently. DIN \cite{din} employs an attention-like module focused on user history, which they call the interest network. Subsequently, Zhou \etal \cite{zhou2019deep} extended DIN to DIEN by incorporating GRU units to capture the temporal evolution of user interests. 

With the increase in the number of users and items and the emergence of various business scenarios, the increasing complexity of user-item interactions places demands on finer-grained predictions.
This leads to the concept of multi-domain CTR (MDCTR) prediction. MDCTR prediction addresses the need to understand and predict user behavior across various domains for more accurate recommendations.

\subsection{Multi-Domain CTR (MDCTR) Prediction}
In many commercial click-through rate prediction scenarios (e.g., domain),  different user groups (e.g., new and old users)~\cite{li2020improving}, different channel modules of APPs~\cite{he2020dadnn, sheng2021one}, different item categories~\cite{zou2022automatic}, etc., can be regarded as different domains. There are obvious differences in the data distribution of users and items in different domains. If each domain is built with an independent model, it may ignore the commonalities among domains~\cite{mtl}, making it difficult to learn from long-tail domains effectively.
Furthermore, training a model for each domain would significantly increase the resource and manpower costs of the system. 
If we directly mix the samples and train a single model, it will overlook the variability of different domains, decreasing prediction accuracy~\cite{mmoe}. In addition, if the data samples are imbalanced across different domains, the model may be dominated by the domains with larger amounts of data, resulting in poor learning performance for the smaller domains. Therefore, adopting MDCTR prediction approaches has become the mainstream solution in the industry~\cite{m2m}.

One of the key challenges in MDCTR is the \textbf{Domain Seesaw Phenomenon}~\cite{ple}. This challenge arises because, in industrial applications, different domains have different amounts of data. Models are highly susceptible to being dominated by domains with large amounts of data, thus decreasing the accuracy of predictions for other domains with less data. 
Most existing MDCTR models typically solve this problem by separating the model into multiple domain-specific parts and shared parts to learn domain commonalities and characteristics individually. 
For example, Ma~\etal introduced the MMOE model \cite{mmoe}, which handles the trade-off between domain commonalities and distinctions using multiple experts in a shared-bottom architecture.
PLE~\cite{ple}, a progressive layered extraction model, adjusting the balance between shared and domain-specific components, allowing for a more nuanced approach to handling inter-domain dynamics. 
While existing models have attempted to address the `Domain Seesaw Phenomenon' through structural modifications, they still struggle significantly in some data-sparse domains~\cite{chang2023pepnet}. In contrast, our \mname leverages the extensive background knowledge of LLMs about the world. This repository of pre-existing knowledge allows our model to handle a wide range of domains, avoiding the seesaw issue. 

Another challenge in MDCTR is the \textbf{under-utilization of semantic information} due to the reliance on sparse feature embedding of mainstream models, leading to a loss of critical domain context and meaning. To address this challenge, various approaches represent strides in addressing the semantic information under-utilization challenge by introducing domain-aware structures and independent embeddings for each domain. DADNN \cite{he2020dadnn} utilizes a domain-aware structure to preserve the unique characteristics of different domains. Tan \etal \cite{tan2021multi} explore the non-shared embeddings for each task and domain, reducing their coupling. However, these attempts still use traditional feature engineering to enhance the remaining features, which does not essentially address the loss of semantic information. Instead, our model concatenates the input features with additional background texts into the textual form of prompts, which are embedded at the token level through a tokenizer and fed into the LLM, which maximizes the preservation of the original input information.

Furthermore, the \textbf{scalability} of MDCTR models is often limited by their architectures and feature embedding methods, which can struggle to adapt to new and evolving domains. Scalability in this context refers to a model's ability to efficiently handle an increasing number of domains or adapt to a new domain without requiring extensive retraining or reconstruction of the entire model.
Some existing methods attempt to improve the model's scalability by separating hard-sharing networks with different purposes \cite{hamur}.
For instance, 
MTMS \cite{tan2021multi} proposes a unified ranking model that integrates independent embedding layers and unified feature management to improve performance and scalability. 
STAR \cite{sheng2021one} adds a separate domain-specific tower for each domain to enable new domains to be joined independently. However, MTMS lacks the ability to model the characteristics of domains. As the number of domains increases, the parameters of STAR models must increase exponentially. Our model overcomes these shortcomings by decoupling LLMs and domain-specific networks to model the commonalities and characteristics of domains, respectively and exhibits good scalability.

\subsection{LLM for CTR Prediction}

Natural language processing (NLP) has witnessed rapid advancements in language model development. These models can be categorized based on their scale, ranging from \textbf{Language Models} (LMs) to \textbf{Large Language Model} (LLMs). Initial advancements are marked by models like BERT \cite{bert} and RoBERTa \cite{liu2019roberta}, which employ pre-training tasks such as Masked Language Model (MLM) and achieve impressive performance in various downstream text comprehension tasks~\cite{plm-mlm}.

As the field progressed, the emergence of LLMs like DeBERTa-large \cite{he2020deberta}, GPT-3 \cite{brown2020language}, and LLaMA \cite{touvron2023llama} represented a huge leap in language modeling. These models, characterized by their enormous size and capacity, demonstrate incomparable proficiency in text generation and understanding human conversations~\cite{llmsurvey}. They have been extensively pre-trained on huge datasets and thus have vast world knowledge that captures deep user demands, making them highly effective in addressing CTR prediction challenges.

Incorporating BERT and its variants in CTR prediction marked a substantial development. Approaches like CTR-BERT \cite{muhamed2021ctr} and DCAF-BERT \cite{muhamed2022dcaf} blended textual and numerical data effectively, demonstrating the potential of language modeling for CTR prediction. Similarly, CTR-BERT~\cite{muhamed2021ctr} enhances the click-through rate prediction by employing a dual-tower structure based on separate user and item language models coupled with a distillation scheme.

The latest trend in CTR prediction is marked by the use of \textbf{generative LLMs} like GPT \cite{brown2020language}, LLaMA~\cite{touvron2023llama}, GLM~\cite{du2022glm}. M6-Rec \cite{cui2022m6} revolutionizes the representation of user behavior by treating it as plain text and introduces an innovative prompt tuning method termed ``option tuning''. CTRL \cite{li2023ctrl} and FLIP~\cite{wang2023flip} challenge the traditional one-hot feature encoding process by leveraging pre-trained language models to assimilate semantic signals and external world knowledge, offering a more nuanced approach. The S\&R Multi-Domain Foundation model \cite{gong2023unified} utilizes LLMs to distill domain-invariant text features from queries and items, enhancing CTR predictions in cold start scenarios within online services. Further extending the capabilities of traditional recommendation models, KAR \cite{xi2023towards} harnesses the power of LLMs for open-world reasoning and factual knowledge extraction and adaptation. Despite significant advancements in utilizing LMs for CTR prediction, the field of multi-domain CTR prediction is still relatively unexplored.

\subsection{Feature Engineering for Non-LLM based Systems.} 

In conventional MDCTR prediction methods that do not use language models, continuous user-item features (e.g., age, income, price) can be directly input into the model since they are numerical in nature.
In contrast, categorical ones (e.g., occupation, gender, brand) typically require one-hot encoding.
For example, the user-item features $\boldsymbol{x}$ (Occupation=\texttt{Doctor}, Title=\texttt{Harmonicas (12 ct)}, Brand=\texttt{Fun Express}, \dots) can be represented as a series of one-hot vectors:
\begin{equation}
\label{onehot}
\small
    \boldsymbol{x} = \underbrace{[1,0,0,\dots,0]}_{\text{Occupation}} \underbrace{[0,1,0,\dots,0]}_{\text{Title}} \underbrace{[0,0,1,\dots,0]}_{\text{Brand}}.
    \dots.
\end{equation}
Conventional approaches will map these one-hot encoded features into a dense vector space through an embedding matrix. However, this process significantly loses crucial semantic information, which is often the key to understanding and distinguishing different domains. 
Consequently, many existing MDCTR models have not adequately considered such critical semantic information, which hinders their ability to model the commonalities and characteristics across multiple domains.

\section{Task Formulation}
\label{sec:task-formulation}

In this section, we introduce the task formulation of multi-domain CTR (MDCTR) prediction . MDCTR aims to build a model that accurately predicts the probability of users clicking on recommended items across various domains. Different user groups (e.g., new and old users)~\cite{li2020improving}, different channel modules of APPs~\cite{he2020dadnn, sheng2021one}, different item categories~\cite{zou2022automatic}, etc., can be regarded as different domains. Given an MDCTR dataset with $M$ distinct domains $\boldsymbol{D} = \{d_1, d_2, \ldots, d_M\}$, $d_m$ represents user-item data from a specific domain, denoted as:
\begin{equation}
\begin{aligned}
  \mathcal{D}_{d_m}
  &= \left \{(\boldsymbol{x}_{i}^{d_m}, y_{i}^{d_m}), i\in \left |\mathcal{D}_{d_m}\right | \right \},
\end{aligned}
\end{equation}
where $\boldsymbol{x}_{i}^{d_m}$ represents the user-item feature set and $y_{i}^{d_m} \in \{0,1\}$ is the click label, with 1 indicating a click and 0 indicating none. $\left |\mathcal{D}_{d_m}\right |$ denotes the total number of samples in the dataset corresponding to domain $d_m$.

For a specific domain $d_m$, when given feature $\boldsymbol{x}_i^{d_m}$ and corresponding binary labels $y_i^{d_m}$,  the model $f(\cdot)$ predicts the click-through rate, denoted by $\hat{y}_i^{d_m}$, which can be represented as : 
\begin{equation}
\begin{aligned}
    \hat{y}_i^{d_m} = P(y_{i}^{d_m} = 1 \mid \boldsymbol{x}_i^{d_m}, d_m) &= f(\boldsymbol{x}_i^{d_m}, d_m).
\end{aligned}
\end{equation}
To align the predictions $\hat{y}_i^{d_m}$ with the actual click labels $y_i^{d_m}$, the binary cross-entropy (BCE) Loss will commonly be used to optimize the model:
\begin{equation}
    \begin{aligned}
    \mathcal{L} 
    & = BCE(y_i^{d_m}, \hat{y}_i^{d_m}) \\
    & = -\left[ y_i^{d_m} \log(\hat{y}_i^{d_m}) + (1 - y_i^{d_m}) \log(1 - \hat{y}_i^{d_m}) \right]. \\
    \end{aligned}
\end{equation}

\section{The Proposed Method}
\label{sec:method}

In this section, we elaborate on our proposed \mname model in detail. First, we provide an overview framework of \mname in Section~\ref{ssec:framework}. Then we describe our prompting strategy for semantic encoding in Section~\ref{ssec:prompt}, followed by the Uni-CTR architecture and further prediction in Section~\ref{ssec:architecture} and Section~\ref{ssec:loss}, respectively. Finally, a horizontal comparison against exiting MDCTR systems is performed in Section~\ref{ssec:discussion} to illustrate the advantage of our model design.

\subsection{Framework Overview}
\label{ssec:framework}

\mname mainly consists of three modules, as illustrated in Fig.~\ref{fig:unictr}. 
Firstly, input data points, including domain, user, and product features, are consolidated into natural language sequences via a predetermined prompt template.
These prompt sequences are fed into an \textbf{LLM backbone} to generate contextualized semantic representations. 
Subsequently, the \textbf{domain-specific networks} leverage the LLM representations from the different transformer layers through ladder networks, thereby learning domain-specific characteristics.
Meanwhile, a \textbf{general network} utilizes the LLM representations from the final transformer layer to model the commonalities across all domains. We demonstrate that the general network can greatly improve the performance of zero-shot prediction on unseen domains. 
Moreover, a \textbf{masked loss strategy} is designed to ensure the decoupling of domain-specific and general networks. Hence, we can incorporate new domains to \mname without affecting other domain-specific networks.

\subsection{Prompt-based Semantic Modeling}
\label{ssec:prompt}

\begin{figure}[h]
    \centering
    \includegraphics[width=\linewidth]{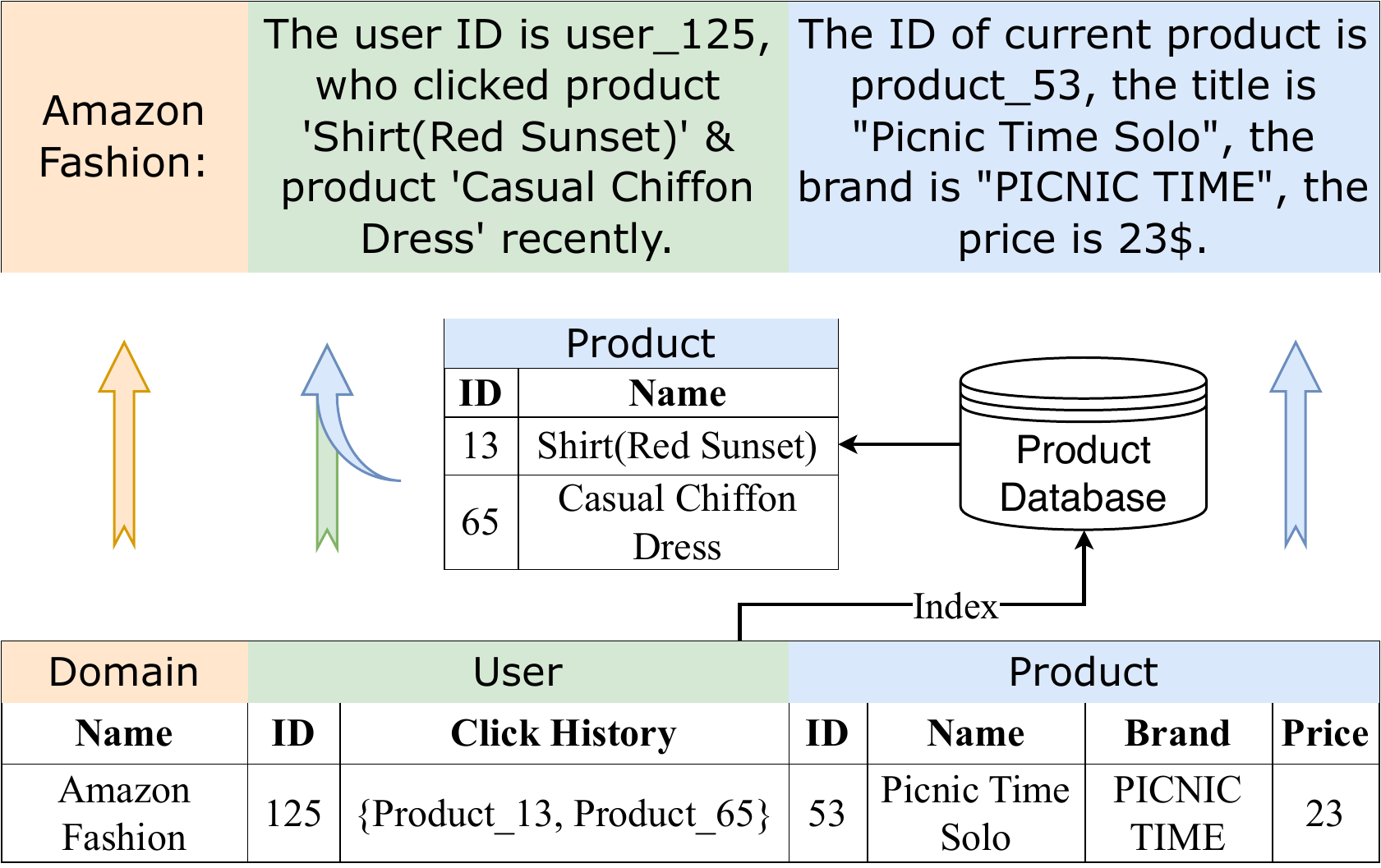}
    \caption{Prompt Template Design to consolidate domain, user, and product features for \mname.}
    \label{fig:prompt}
\end{figure}

We design a prompt-based modeling approach to capture the rich semantic information of text-based features, rather simply encoding them as one-hot vectors in traditional approaches \cite{man2017cross, hu2018conet, zhang2023collaborative, chen2023cross}.
As demonstrated in Fig.~\ref{fig:prompt}, our prompt incorporates feature information from the following three sources: 
\begin{itemize}[leftmargin=*]
\item \textbf{Domain Context ($d$)} preserves domain information by explicitly appending the domain type at the start of the prompt sequence. This component enables \mname to comprehend the input domain, thereby effectively learning and distinguishing features between different domains. 
\item \textbf{User Information ($u$)} captures users' behavioral patterns and preferences, including user IDs and click history. Click history lists the IDs of products recently viewed by the user. Note that we substitute product IDs with their corresponding textual descriptions extracted from the product database.
\item \textbf{Product Information ($p$)} exhibits extensive product description, including the unique product ID, title, brand, and price. 
\end{itemize}

Consequently, the input data can be represented by the following aggregation: $x = \left \{ d, u, p \right \}$, representing domain, user, and product feature, respectively. We design the following prompt template to consolidate these features into textual sequences $x_{\text{text}}$:
\begin{equation}
\label{eq:prompt}
\begin{aligned}
    x_{\text{text}} = \, & \text{[Domain Name]: The user ID is user\_} \\
    & \text{[User ID], who clicked product `[Product1 Title]'} \\
    & \text{and product `[Product2 Title]' recently.} \\
    & \text{The ID of the current product is product\_} \\
    & \text{[Product ID], the title is [Product Name],} \\
    & \text{the brand is [Brand], the price is [Price].} \\
\end{aligned}
\end{equation}
where [Product Title], [Product ID],  [Brand], and [Price] are substituted by information retrieved from the product database.

Using our prompt-based semantic modeling design, \mname is able to preserve all semantic features in the prompt format (textual sequence). Our LLM backbone can subsequently utilize these prompts to capture both commonalities and characteristics of multiple domains.

\subsection{\mname architecture} 
\label{ssec:architecture}

\begin{figure*}[h]
  \centering
  \includegraphics[width=\linewidth]{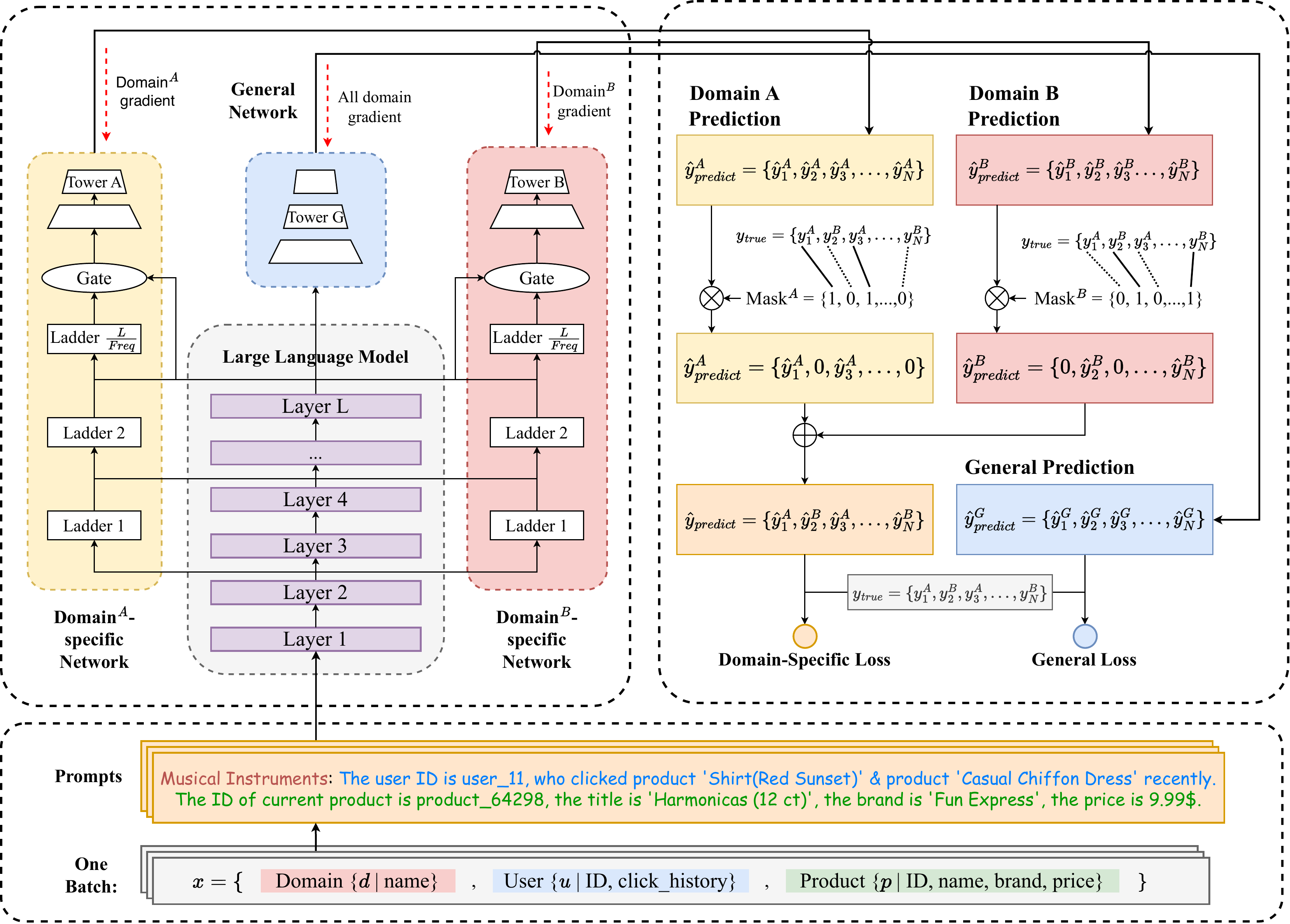}
  \caption{The architecture of the \mname, which takes the prompts as the input and obtains semantic representations using an LLM. Among them, domain-specific networks address the characteristics of each domain, while an additional general network aims to extract the commonalities among domains.}
  \label{fig:unictr}
\end{figure*}

As shown in Fig.~\ref{fig:unictr}, Uni-CTR mainly comprises three components: \textbf{LLM Backbone} (Section~\ref{ssec:llm-backbone}), \textbf{Domain-Specific Network} (Section~\ref{ssec:dsn}), and \textbf{General Network} (Section~\ref{ssec:gen_network}). 

\subsubsection{\textbf{LLM Backbone}}
\label{ssec:llm-backbone}

As the backbone of \mname, LLM serves to encode the rich semantic contextual information from the input sequence $x_{\text{text}}$ (see in Equation~\ref{eq:prompt}).

The prompt $x_{\text{text}}$ is first tokenized into a sequence of tokens $t_j$:
\begin{equation}
    \boldsymbol{x}_{\text{tokens}} = \text{Tokenizer}(x_{\text{text}}) = \{t_0, t_1, \ldots, t_J\}, 
\end{equation}
where each token $t$ corresponds to a unique ID in a predefined token dictionary.

Next, these tokens are passed into the LLM embedding layer $\boldsymbol{E}_{embed}$, where each token $t_j$ is mapped to a fixed-dimensional vector $ \boldsymbol{e}_j $, which contains the semantic and positional information. 
The initialized representation is written by:
\begin{equation}
    \boldsymbol{h}_0 = \boldsymbol{E}_{embed}(\boldsymbol{x}_{\text{tokens}}) = \{\boldsymbol{e}_0, \boldsymbol{e}_1, \ldots, \boldsymbol{e}_J\}.
\end{equation}

Following the embedding layer, $\boldsymbol{h}_0$ is passed to the LLM's encoder or decoder, which typically consists of multiple transformer layers.
Each layer $l$ of the encoder or decoder receives the output $\boldsymbol{h}_{l-1}$ from its preceding layer and produces a new output $\boldsymbol{h}_l$, which can be defined as follows:
\begin{equation}
    \boldsymbol{h}_l = \text{Transformer}_l(\boldsymbol{h}_{l-1}), l\in \{ 1,2,\ldots, L\},
\end{equation}
where we assume that the encoder or decoder of the LLM has $\boldsymbol{L}$ layers. 
We utilize the representations from the embedding layer and all $\boldsymbol{L}$ transformer layers. The collection of all representations $\boldsymbol{H}$ is described as: 
\begin{equation} \label{eq:h}
    \boldsymbol{H} = \{\boldsymbol{h}_0, \boldsymbol{h}_1, \boldsymbol{h}_2, \ldots, \boldsymbol{h}_L\},
\end{equation}
The semantic representations $\boldsymbol{H}$ generated by the LLM backbone can capture the commonalities and coarse-grained characteristics across all domains. In order to model more fine-grained domain characteristics, $\boldsymbol{H}$ is fed into domain-specific networks (see Section~\ref{ssec:dsn}) for characteristics of distinct domains.

\subsubsection{\textbf{Domain-Specific Network}}
\label{ssec:dsn}

Traditional models often struggle to adapt to continuously evolving user-item features, particularly when integrating new domains. This challenge typically requires significant modifications and retraining of models. To overcome this limitation, we design the Domain-Specific Network (DSN), a modular module that laterally interfaces with the LLM backbone, akin to a ladder.
As depicted in Fig.~\ref{fig:unictr}, each domain corresponds to a DSN to extract its fine-grained and domain-specific features. Notably, fine-tuning existing DSN or traning a new DSN does not influence other DSNs.
The DSN comprises three principal components: 
(i) the \textbf{ladder network} to extract intermediate representations from the LLM backbone,  
(ii) the \textbf{gate network} to regulate the information that passes through the ladder network,
and (iii) the \textbf{tower network} to make in-domain predictions.

\paragraph{\textbf{Ladder Network}}

Existing MDCTR methods such as STAR~\cite{sheng2021one} train multiple domain-specific networks that mirror the size and structure of the backbone network.
In contrast, we train a ladder side network~\cite{sung2022lst}, a small and independent network that receives intermediate representations $\boldsymbol{H}$ (in Equation~\ref{eq:h}) through shortcut connections, referred to as ladders, from the backbone LLM.

The number of ladders is proportionally correlated with the number of representations within $\boldsymbol{H}$, as obtained from the LLM backbone. 
However, introducing one ladder for every component of $\boldsymbol{H}$ would lead to an excessive number of parameters for domain-specific networks. To reduce the size of the ladder network, we define a frequency hyper-parameter $\phi$ (denoted by `\textit{Freq}' in Fig.~\ref{fig:unictr}), and deploy one ladder for every $\phi$ transformer layers of the backbone LLM.
Given a ladder network with $F=\frac{L}{\phi}$ ladders, the output $\boldsymbol{lad}_{f}$ of the $f_{\text{th}}$ ladder is calculated as follows:
\begin{equation} \label{eq:ladder_f}
\begin{aligned}
    \boldsymbol{lad}_{f} = 
    \begin{cases} 
        Ladder_{1}(\boldsymbol{h}_\phi) & \text{if } f = 1 \\
        Ladder_{f}(\boldsymbol{h}_{f \cdot \phi} + \boldsymbol{lad}_{f-1}) & \text{if } f \in \{ 2,\ldots, F\} \\
    \end{cases},
\end{aligned}
\end{equation}
where $Ladder_{f}$ can be Multilayer Perceptrons (MLP) \cite{murtagh1991multilayer}, Attention Networks \cite{visualattention}, Transformer Blocks (with self attention) \cite{vaswani2017attention}, or any other types of neural networks. 
Note that $Ladder_{1}$ takes only the $\phi_{\text{th}}$ representations $\boldsymbol{h}_\phi$ from the LLM backbone as the input, whereas subsequent $Ladder_{f}$ relies on the representation $\boldsymbol{h}_{f \cdot \phi}$ as well as the output from previous ladder.

\paragraph{\textbf{Gate Network}}

For each domain, the target features to be learned comprise (i) common features shared across all domains and (ii) domain-specific features. Specifically, the ladder network's output captures the domain-specific characteristics, while the final hidden states of the LLM backbone capture the commonalities across all domains. Therefore, each domain's predictions need to leverage the ladder and LLM backbone outputs. To facilitate this integration, we design a gate network that follows the ladder network. The gate network is based on a dynamic weight assignment mechanism, aiming to adaptively balance domain-specific and common features.

Firstly, we concatenate the outputs of the last ladder and the final transformer layer of the LLM backbone:
\begin{equation}
\label{eq:ladder_o}
    \boldsymbol{O} = \text{concat}(\boldsymbol{h}_L, \boldsymbol{lad}_{F}).
\end{equation}

Subsequently, we utilize an attention pooling~\cite{attention-pooling} module to compute the attention weights across the concatenated features in Equation~\ref{eq:ladder_o}. This is to adjust the propositions of the domain-specific and common features dynamically:
\begin{equation}
\begin{aligned}
    \boldsymbol{score} &= \textrm{tanh}(\boldsymbol{W}_{k}\boldsymbol{O}) \boldsymbol{W}_{q}, \\
    \boldsymbol{A} &= \textrm{softmax}(\boldsymbol{score}), \\
    \boldsymbol{R} &= \boldsymbol{A}^{T}\boldsymbol{O},
\end{aligned}
\end{equation}
where $\boldsymbol{W}_{k}$ and $\boldsymbol{W}_{q}$ are the learnable parameters for computing the key and query, $ \boldsymbol{A}$ is the attention weights normalized from $\boldsymbol{score}$ by a softmax operation.
Lastly, $\boldsymbol{R}$ represents the pooled weights, including common and specific features of a certain domain.
Therefore, for each domain $d_m$, the corresponding gate network produces a compressed representation, denoted by $\boldsymbol{R}^{d_m}$.

\paragraph{\textbf{Tower Network}}

The last module of the DSN, the tower network, will use $\boldsymbol{R}^{d_m}$ to make final predictions $\boldsymbol{\hat{y}}^{d_m}$ (or $\boldsymbol{\hat{y}}^{d_m}_{\text{predict}}$ in Fig.~\ref{fig:unictr}), where the ground truth label is $\boldsymbol{y}^{d_m}$ (or denoted as $\boldsymbol{y}^{d_m}_{\text{true}}$ in Fig.~\ref{fig:unictr}). Each domain-specific network has one tower network, which is an MLP network as follows:
\begin{equation}
\begin{aligned}
    \boldsymbol{\hat{y}}^{d_m} &= \text{MLP}(\boldsymbol{R}^{d_m}; \boldsymbol{W}_{\gamma}^{d_m}, \boldsymbol{b}_{\gamma}^{d_m}),
\end{aligned}
\end{equation}
where $\boldsymbol{W}_{\gamma}^{d_m}$ and $\boldsymbol{b}_{\gamma}^{d_m}$ denote the weights and biases of the MLP network. $\boldsymbol{\hat{y}}^{d_m}$ is the prediction of whether the user will click it on this item or not.

\subsubsection{\textbf{General Network}}
\label{ssec:gen_network}

In the previous Section~\ref{ssec:dsn}, we design distinct DSNs to model and predict within specific seen domains.
For \mname to operate effectively with new, unseen domains, we have constructed a general network that captures the commonalities of all established domains.
Specifically, the general network contains only a tower network and directly utilizes the last hidden states ${h}_L$ of the LLM backbone via MLP network, which is defined as follows:
\begin{equation}
\begin{aligned}
    \boldsymbol{\hat{y}}^{G} &= \text{MLP}(\boldsymbol{h}_L; \boldsymbol{W}_{\sigma}^{G}, \boldsymbol{b}_{\sigma}^{G}),
\end{aligned}
\end{equation}
where $\boldsymbol{W}_{\sigma}^{d_m}$ and $\boldsymbol{b}_{\sigma}^{d_m}$ denote the weights and biases of the MLP network.

The general network can adapt to situations where no prior data exists, i.e., the data-sparse domains, serving as a zero-shot predictive model. In cold-start recommendation scenarios, this capability is crucial because it allows the network to maintain predictive accuracy without the need to target any domain-specific training data.

\subsection{Masked Loss Strategy}
\label{ssec:loss}

As analyzed in Section~\ref{ssec:dsn}, the DSNs compute predictions for all domain samples during the training phase.
To avoid redundant predictions being taken into account by the loss function, which in turn makes the update of the DSN parameters affected by other domains, we use a masked loss strategy. 

\subsubsection{Masked Multi-domain Prediction}

The prediction of \mname is based on either a DSN for a known domain or the General Network for an unknown domain. For a data sample from known domains ($d_m \in \boldsymbol{D}$), the predictions of all DSNs is denoted as $\boldsymbol{\hat{y}}^{D} = [ \hat{y}^{d_1}, \hat{y}^{d_2},\ldots,\hat{y}^{d_M} ]$. 
Note that we have $M$ number of DSNs if there are $M$ domains.
To remove irrelevant predictions, we keep only the prediction of one DSN whose domain is the same as $d_m$ via a mask. The mask is generated as follows:
\begin{equation} \label{eq:mask}
    \boldsymbol{mask}^{d_m} = \left [ I(d_1=d_{m}), I(d_2=d_{m}),\ldots,I(d_M=d_{m})\right ], 
\end{equation}

where $I(\cdot)$ is an indicator function that equals $1$ if the conditional statement is true and $0$ otherwise.
Given a data sample from an unknown domain $d_m \notin \boldsymbol{D}$, the prediction of \mname is determined by the output $\hat{y}^{G}$ of the general network. Accordingly, the prediction of \mname can be formulated as follows:
\begin{equation} \label{eq:prediction}
\begin{aligned}
    \hat{y} = 
    \begin{cases} 
        \text{sum} \left ( \boldsymbol{mask}^{d_m} \cdot \boldsymbol{\hat{y}}^{D} \right ) & \text{if } d_m \in \boldsymbol{D} \\
        \hat{y}^{G} & \text{if } d_m \notin \boldsymbol{D} \\
    \end{cases}.
\end{aligned}
\end{equation}

\subsubsection{Masked Multi-domain Training}

\paragraph{\textbf{Loss Computation}} The loss of each data sample comprises of: (i) domain-specific loss $\mathcal{L}^{D}$ calculated by the prediction of the domain $d_m$'s DSN, and (ii) general loss $\mathcal{L}^{\text{G}}$ computed based on the general network's prediction.
The overall loss can be written as:
\begin{equation}
    \mathcal{L} = \mathcal{L}^{D} + \mathcal{L}^{G}
\end{equation}
where the domain-specific loss $\mathcal{L}^{D}$ can be computed as follows:
\begin{equation}
    \begin{aligned}
        \mathcal{L}^{D}
        & = \text{sum} \left ( \boldsymbol{mask}^{d_m} \odot \left [ \ell( \hat{y}^{d_1}, y), \ell( \hat{y}^{d_2}, y), \ldots, \ell( \hat{y}^{d_M}, y) \right ] \right ) \\
        & = \sum_{i=1}^{M} \left( mask^{d_m}_i \cdot \ell(\hat{y}^{d_i}, y) \right) \\
        & = mask^{d_m}_m \cdot \ell(\hat{y}^{d_m}, y) \\
        & = \ell(\hat{y}^{d_m}, y),
    \end{aligned}
\end{equation}
where $y$ is the ground truth label of the data sample, $\ell(\cdot)$ denotes the Binary Cross-Entropy Loss (BCELoss). The $mask^{d_m}_i$ is the $i$-th binary element $I(d_i=d_{m})$ within $\boldsymbol{mask}^{d_m}$.
And the general loss $\mathcal{L}^{\text{G}}$ can be calculated as:
\begin{equation}
    \mathcal{L}^{G} = \ell(\hat{y}_{G}, y)
\end{equation}

\paragraph{\textbf{Loss Back-propagation}}

The trainable components of \mname are the LLM backbone, DSNs, and the general network. 
We denote the parameters of them to be $\boldsymbol{\theta}_{\text{LLM}}$, $\boldsymbol{\theta}_{\text{DSN}}$ and $\boldsymbol{\theta}_{\text{G}}$, respectively.
Given a data sample $(x^{d_m}, y^{d_m})$ from domain $d_m$, we use the following loss to update the parameters of \mname:
\begin{equation}
    \begin{aligned}
        \mathcal{L} = \mathcal{L}^{D} \left ( x^{d_m} ; \boldsymbol{\theta}_{\text{LLM}} , \boldsymbol{\theta}_{\text{DSN}} \right ) 
        + \mathcal{L}^{G} \left ( x^{d_m};  \boldsymbol{\theta}_{\text{LLM}}, \boldsymbol{\theta}_{\text{G}} \right )
    \end{aligned}
\end{equation}

The gradient with respect to the LLM backbone parameters $\boldsymbol{\theta}_{\text{LLM}}$ is calculated as follows:
\begin{equation}
    \begin{aligned}
        \nabla_{\boldsymbol{\theta}_{\text{LLM}}} \mathcal{L} 
        & = \nabla_{\boldsymbol{\theta}_{\text{LLM}}}\mathcal{L}^{D} \left ( x^{d_m} ; \boldsymbol{\theta}_{\text{LLM}} , \boldsymbol{\theta}_{\text{DSN}} \right ) \\
        & + \nabla_{\boldsymbol{\theta}_{\text{LLM}}} \mathcal{L}^{G} \left ( x^{d_m};  \boldsymbol{\theta}_{\text{LLM}}, \boldsymbol{\theta}_{\text{G}} \right ) \\
        & = \frac{\partial\mathcal{L}^{D} \left ( x^{d_m} ; \boldsymbol{\theta}_{\text{LLM}} , \boldsymbol{\theta}_{\text{DSN}} \right )}{\partial \boldsymbol{\theta}_{\text{LLM}}}  + \frac{\partial \mathcal{L}^{G} \left ( x^{d_m};  \boldsymbol{\theta}_{\text{LLM}}, \boldsymbol{\theta}_{\text{G}} \right )}{\partial \boldsymbol{\theta}_{\text{LLM}}} \\
        & = \frac{\partial \boldsymbol{mask}^{d_m} \odot \left [ \ell( \hat{y}^{d_i}, y ; \boldsymbol{\theta}_{\text{LLM}} , \boldsymbol{\theta}_{\text{DSN}}^{d_i} )  , i \in [M] \right ]}{\partial \boldsymbol{\theta}_{\text{LLM}}}  \\
        & + \frac{\partial \ell \left ( \hat{y}_{G}, y;  \boldsymbol{\theta}_{\text{LLM}}, \boldsymbol{\theta}_{\text{G}} \right )}{\partial \boldsymbol{\theta}_{\text{LLM}}}  \\
        & = \frac{\partial  \ell( \hat{y}^{d_m}, y ; \boldsymbol{\theta}_{\text{LLM}} , \boldsymbol{\theta}_{\text{DSN}}^{d_m} )}{\partial \boldsymbol{\theta}_{\text{LLM}}}  + \frac{\partial \ell \left ( \hat{y}_{G}, y;  \boldsymbol{\theta}_{\text{LLM}}, \boldsymbol{\theta}_{\text{G}} \right )}{\partial \boldsymbol{\theta}_{\text{LLM}}}. \\
    \end{aligned}
\end{equation}

We then elaborate the gradient for DSN parameters $\boldsymbol{\theta}_{\text{DSN}}$. Note that $\boldsymbol{\theta}_{\text{DSN}} = [\boldsymbol{\theta}_{\text{DSN}}^{d_1}, \ldots, \boldsymbol{\theta}_{\text{DSN}}^{d_M} ]$ consists of $M$ sets of DSN parameters. The gradient of each DSN of domain $d_n, n \in \{ 1, 2, \ldots, M \} $ is is calculated as follows:
\begin{equation}
    \begin{aligned}
        \nabla_{\boldsymbol{\theta}_{\text{DSN}}^{d_n}} \mathcal{L} 
        & = \nabla_{\boldsymbol{\theta}_{\text{DSN}}^{d_n}}\mathcal{L}^{D} \left ( x^{d_m} ; \boldsymbol{\theta}_{\text{LLM}} , \boldsymbol{\theta}_{\text{DSN}} \right ) \\
        & + \nabla_{\boldsymbol{\theta}_{\text{DSN}}^{d_n}} \mathcal{L}^{G} \left ( x^{d_m};  \boldsymbol{\theta}_{\text{LLM}}, \boldsymbol{\theta}_{\text{G}} \right ) \\
        & = \frac{\partial\mathcal{L}^{D} \left ( x^{d_m} ; \boldsymbol{\theta}_{\text{LLM}} , \boldsymbol{\theta}_{\text{DSN}} \right )}{\partial \boldsymbol{\theta}_{\text{DSN}}^{d_n}} \\
        & = \frac{\partial \boldsymbol{mask}^{d_m} \odot \left [ \ell( \hat{y}^{d_i}, y ; \boldsymbol{\theta}_{\text{LLM}} , \boldsymbol{\theta}_{\text{DSN}}^{d_i} ), i \in [M] \right ]}{\partial \boldsymbol{\theta}_{\text{DSN}}^{d_n}}  \\
        = & \begin{cases} 
                0 & \text{if } n \ne m \\
                \frac{\partial  \ell( \hat{y}^{d_m}, y ; \boldsymbol{\theta}_{\text{LLM}} , \boldsymbol{\theta}_{\text{DSN}}^{d_m} )}{\partial \boldsymbol{\theta}_{\text{DSN}}^{d_m}}. & \text{if } n = m\\
            \end{cases},
    \end{aligned}
\end{equation}
which indicates that only parameters of DSN $d_m$ can be updated due to the mask we defined in Equation~\ref{eq:mask}. The parameters of all DSNs other than $d_m$ remain unchanged, ensuring the decoupling of gradient updates across various domains. Therefore, each DSN can independently model the characteristics of each domain without affecting other DSNs.

Finally, the gradient of the loss with respect to the general network parameters $\boldsymbol{\theta}_{\text{G}}$ is computed as follows:
\begin{equation}
    \begin{aligned}
    \nabla_{\boldsymbol{\theta}_{\text{G}}} \mathcal{L} 
    & = \nabla_{\boldsymbol{\theta}_{\text{G}}}\mathcal{L}^{D} \left ( x^{d_m} ; \boldsymbol{\theta}_{\text{LLM}} , \boldsymbol{\theta}_{\text{DSN}} \right ) + \nabla_{\boldsymbol{\theta}_{\text{G}}} \mathcal{L}^{G} \left ( x^{d_m};  \boldsymbol{\theta}_{\text{LLM}}, \boldsymbol{\theta}_{\text{G}} \right ) \\
    & = \frac{\partial \mathcal{L}^{G} \left ( x^{d_m};  \boldsymbol{\theta}_{\text{LLM}}, \boldsymbol{\theta}_{\text{G}} \right )}{\partial \boldsymbol{\theta}_{\text{G}}} \\
    & = \frac{\partial \ell \left ( \hat{y}_{G}, y;  \boldsymbol{\theta}_{\text{LLM}}, \boldsymbol{\theta}_{\text{G}} \right )}{\partial \boldsymbol{\theta}_{\text{G}}}.
    \end{aligned}
\end{equation}
This gradient ensures that the general network can be trained based on data samples from all domains. Hence, the general network can effectively learn different cross-domain features, and thereby, it can be generalized to unseen domains for accurate zero shot prediction.

In summary, our proposed masked loss strategy ensures that the domain-specific networks are effectively tailored to their respective tasks, while the LLM backbone and the general network learn the domains' commonalities. This architecture enables (i) DSNs to be pluggable and scalable and make accurate predictions on known domains and (ii) the general network to robustly predict unseen domain samples.

\subsection{Horizontal Comparison and Discussion}
\label{ssec:discussion}

In this section, we conduct a comparative analysis of \mname against existing MDCTR systems (as summarized in Table~\ref{tab:discussion}), emphasizing the advancement and contribution of our proposed method.

\begin{table}[htbp]
    \centering
    \caption{Comparison of our methodology with several previous research studies. ``Semantics'' represents the ability of the model to deal with complex semantics. ``Balance'' represents the ability of the model to deal with see-saw problems. ``Generalization'' represents the ability of the model to deal with unknown new domains. ``Scalability'' represents the scalability of the model's structure.}
    \resizebox{\linewidth}{!}{%
        \begin{tabular}{@{}ccccc@{}} 
            \toprule
            MDCTR Model    & Semantics & Balance & Generalization & Scalability \\ \midrule
            Shared Bottom~\cite{sharebottom}    &  $\times$   & $\times$  & $\times$  & $\checkmark$  \\ 
            MMOE~\cite{mmoe}, PLE~\cite{ple} &   $\times$       &   $\checkmark$    &  $\times$           &  $\times$            \\
            STAR~\cite{sheng2021one}    &   $\times$        & $\checkmark$         & $\times$         &  $\checkmark$           \\
            \mname  & $\checkmark$         &    $\checkmark$      &  $\checkmark$          &  $\checkmark$           \\ \bottomrule
        \end{tabular}
        \label{tab:discussion}
    }
\end{table} 

The Shared Bottom model~\cite{sharebottom} utilizes a neural network as the shared bottom layer to capture the common information across different domains while incorporating multiple expert networks in the upper layers to model the specific characteristics of each domain. The structure of \mname appears similar to the shared bottom structure. However, \mname differs in two aspects. Firstly, \mname integrates diverse semantic representations from different layers of the LLM backbone, whereas Shared Bottom only utilizes the representation from the final layer. Secondly, \mname incorporates a General Network, enabling it to model shared information across all domains. This feature empowers \mname to be a zero-shot predictor and perform well in unknown domains.

Both MMOE~\cite{mmoe} and PLE~\cite{ple} models target to address the imbalance between different domains. They use multiple experts and gate networks to alleviate the seesaw phenomenon. 
However, their components are tightly coupled, making them hard to scale. Specifically, a new model needs to be built from scratch when incorporating a new domain.
Similarly, when a domain becomes obsolete due to business changes, removing the corresponding expert network and tower structure is challenging due to this coupling. 
In comparison, all DSNs of \mname are pluggable and decoupled, allowing the model to incorporate new or remove existing domains easily.

The parameter-sharing mechanism in STAR~\cite{sheng2021one} offers a solution with fundamental scalability and balance capabilities. However, it neither understands complex semantics nor works under unseen domains.
Moreover, the element-wise computation in STAR requires all DSNs to mirror the size and structure of the shared network.
Therefore, STAR can suffer from significant structural latency and complexity when applied to many domains.
On the contrary, the parameters of each DSN are significantly less than the parameters of the shared networks in \mname.
The lightweight design of \mname's DSN makes it easier to be trained, and faster in inference.

\section{Experiments}
\label{sec:experiments}

In this section, we first elaborate our experimental settings (\eg datasets, evaluation metrics, baselines, and implementation details) in Section~\ref{ssec:experimental-settings}.
Then, we evaluate and discuss of performance and efficiency of \mname framework against baselines from different perspectives, denoted by six research questions (RQ). In Section~\ref{ssec:performance-comparison} (RQ1), we present the main results of the performance of \mname and baseline models. 
Followed by Section~\ref{ssec:zero-shot-prediction} (RQ2) to elaborate \mname's capability in zero-shot prediction under unseen domains.
Section~\ref{ssec:scale-up} (RQ3) studies the scaling law issue, investigating the impact of using different LLMs of varying size as the backbone of \mname. 
The scalability of \mname to incorporate a new domain is illustrated in Section~\ref{ssec:scalability} (RQ4).
We visualize the representations of \mname in Section~\ref{ssec:visualization} (RQ5), to show the different roles of LLM and DSN in the latent space.
In Section~\ref{ssec:ablation} (RQ6), we perform ablation studies on (i) different combinations of semantic features as input, and (ii) different modules of \mname framework.
In summary, we aim to address the following RQs:

\begin{itemize}[leftmargin=*]
    \item RQ1 ($\S$~\ref{ssec:performance-comparison}):
    Can \mname outperform existing SOTA baselines in MDCTR prediction across multiple domains?
    
    \item RQ2 ($\S$~\ref{ssec:zero-shot-prediction}): Can \mname operate effectively in unseen domains and make accurate predictions in a zero-shot manner?
    
    \item RQ3 ($\S$~\ref{ssec:scale-up}): Does the scaling law regarding model size apply to \mname? Does a larger LLM guarantee better performance?

    \item RQ4 ($\S$~\ref{ssec:scalability}): Can \mname scale to incorporate a new domain by simply adding and fine-tuning a new domain-specific network with other parts frozen?

    \item RQ5 ($\S$~\ref{ssec:visualization}): What are the functionalities of the LLM and DSNs, and their impacts on overall model performance?

    \item RQ6 ($\S$~\ref{ssec:ablation}): How do the main components of \mname, \ie semantic modeling prompt, LLM backbone, and DSNs, contribute to \mname's overall performance?
\end{itemize}

\subsection{Experimental Settings}
\label{ssec:experimental-settings}

\subsubsection{Datasets}
\label{sssec:datasets}

we utilize the \textbf{Amazon Review Data (2018)}~\cite{ni2019justifying}, a widely used dataset in CTR prediction. 
It contains records of user interactions with items on the Amazon shopping site from 1996 to 2008. 
Following previous work~\cite{li2023ctrl,wang2023flip}, we select five categories (Fashion, Digital Music, Musical Instruments, Gift Cards, All Beauty) of products as distinct domains for our experiments.
The statistics of the number of users, products, and data samples are shown in Table~\ref{tab:statistics_amazon}. 
In our experiment, we utilize features including domain names, user IDs, user click history, product IDs, product names, descriptions, and prices. 
We take the items with ratings above $3$ as positive examples and those with ratings equal to or below $3$ as negative examples.
We follow previous works \cite{inttower,li2023ctrl,wang2023flip} to split the 80\%, 10\%, and 10\% of the dataset for training, validation, and testing, respectively.
In classical multi-domain datasets such as 
Ali-CCP\footnote{\url{https://tianchi.aliyun.com/dataset/408}} \cite{ma2018entire}
and 
Ali-Mama\footnote{\url{https://tianchi.aliyun.com/dataset/56}} \cite{gai2017learning, zhou2018deep}, features are anonymized.
Specifically, all feature values are labeled as IDs, with related natural languages masked.
Since LLM-based recommendation relies heavily on the semantic features extracted by the LLM backbone, those ID-based features are unsuitable in our experiment setting.
Thus, we exclude Ali-CCP and Ali-Mama from our experiments.

\begin{table}[htbp]
    \centering
    \caption{Statistics of Amazon Review dataset.}
    \label{tab:statistics_amazon}
    \normalsize
    \begin{tabular}{@{}cccc@{}}
        \toprule
        Domains & Users & Products & Samples \\ 
        \midrule
        Fashion  & 749,233 & 186,189 & 883,636 \\
        Digital Music  & 127,174 & 66,010 & 1,584,082 \\
        Musical Instruments  & 903,060 & 112,132 & 1,512,530 \\ 
        Gift Cards  & 128,873 & 1,547 & 147,194 \\ 
        All Beauty  & 319,335 & 32,486 & 371,345 \\ 
        \bottomrule
    \end{tabular}
\end{table}

\subsubsection{Evaluation Metrics}
\label{sssec:metrics}

To evaluate \mname and other baselines, we use the area under the ROC curve (\textbf{AUC}) metric. The ROC (Receiver Operating Characteristic) plots the True Positive Rate (TPR) against the False Positive Rate (FPR) at various threshold settings. 
The AUC quantifies the overall ability of the model to discriminate between the positive and negative classes in all confidence levels.
Mathematically, it is represented as:
\begin{equation}
    \label{e2}
    \begin{aligned}
        AUC = \int_{0}^{1} TPR(FPR^{-1}(u)) du,
    \end{aligned}
\end{equation}
where $TPR$ is the True Positive Rate, $FPR^{-1}$ is the inverse of the False Positive Rate, and $u$ is the threshold. An AUC value of 1 indicates perfect discrimination, while a value of 0.5 indicates that the performance is equivalent to a random guess. Higher AUC values indicate better performance of the model.

Additionally, we follow previous work~\cite{yan2014coupled,li2023ctrl,din} using \textbf{RelaImpr} metric to quantify the relative performance improvement over the baseline models.
The RelaImpr is calculated as:
\begin{equation}
    \label{e1}
    \begin{aligned}
        RelaImpr = (\frac{AUC(measure\ model)-0.5}{AUC(base\ model)-0.5} -1)\times 100\%.
    \end{aligned}
\end{equation}

\subsubsection{Baselines}
\label{sssec:baselines}

To rigorously assess the performance of \mname, we compare it against a diverse array of baseline models. These models are mainly categorized into two groups: Single-domain and Multi-domain CTR models.

\paragraph{\textbf{Single-Domain Models}}

\begin{itemize}[leftmargin=*]

\item \textbf{PNN.} PNN \cite{qu2016product} introduces a product layer, which employs product operations to capture interactions across different feature categories.

\item \textbf{DCN.} Deep \& Cross Network \cite{wang2017deep} proposes a cross-network explicitly tailored to model bounded-degree feature interactions balancing expressiveness with computational efficiency.

\item \textbf{DeepFM.} DeepFM \cite{guo2017deepfm} integrates Factorization Machines (FM) for low-order feature interactions with deep neural networks for high-order feature interactions.

\item \textbf{xDeepFM.} xDeepFM \cite{lian2018xdeepfm} innovates by introducing the Compressed Interaction Network (CIN) to capture high-order feature interactions efficiently.

\item \textbf{DIEN.} The Deep Interest Evolution Network (DIEN) \cite{zhou2019deep} introduces a two-layer structure, the attention-based Interest Extractor Layer and the Interest Evolving Layer, to capture and evolve user interests over time.

\item \textbf{AutoInt.} AutoInt \cite{song2019autoint} uses multi-head self-attention networks to capture hierarchical input feature interactions.

\item \textbf{FiBiNET.} FiBiNET \cite{fibinet} integrates the Squeeze-Excitation network (SENET) mechanism to assess feature importance dynamically and employs a bilinear function to capture complex feature interactions.

\item \textbf{IntTower.} IntTower~\cite{inttower} is a two-tower model for pre-ranking systems, enhancing interaction between user and item features to improve prediction accuracy and inference efficiency in large-scale applications.

\end{itemize}

\begin{table*}[t]
    \caption{Performance comparison of different models. The boldface denotes the highest score, and the underline indicates the best result of all baselines. $\star$ represents significance level $p$-value $<0.05$ of comparing \mname (BackBone Sheared-LLama) with the best baselines.}
    \centering
    \normalsize
    \label{tab:performance}
    \begin{tabular}{@{}cccccccc@{}}
    \toprule
    \multirow{2}{*}{Category} & \multirow{2}{*}{Models} & \multicolumn{2}{c}{Fashion} & \multicolumn{2}{c}{Musical Instruments} & \multicolumn{2}{c}{Gift Cards} \\ 
    \cmidrule(l){3-8} 
                                   &     & AUC & RelaImpr & AUC & RelaImpr & AUC & RelaImpr \\ 
    \midrule
    \multirow{7}{*}{Single-domain} & PNN & 0.6979 & 27.49\% & 0.6859 & 38.19\% & 0.5959 & 134.20\% \\
                                   & DCN & 0.6985 & 27.10\% & \underline{0.6893} & 35.71\% & 0.6126 & 99.47\% \\
                                   & DeepFM & 0.6982 & 27.30\% & 0.6880 & 36.65\% & 0.5937 & 139.70\% \\
                                   & xDeepFM & \underline{0.7031} & 24.22\% & 0.6892 & 35.78\% & 0.6121 & 100.36\% \\
                                   & DIEN & 0.6995 & 26.47\% & 0.6881 & 36.58\% & 0.6105 & 103.26\% \\
                                   & AutoInt & 0.7003 & 25.96\% & 0.6888 & 36.07\% & 0.5976 & 130.12\% \\
                                   & FiBiNET & 0.6770 & 42.54\% & 0.6878 & 36.79\% & 0.6120 & 100.54\% \\
                                   & IntTower & 0.6988 & 26.91\% & 0.6888 & 36.07\% & 0.6100 & 104.18\% \\ 
    \midrule
    \multirow{4}{*}{Multi-domain} & Shared Bottom & 0.6946 & 29.65\% & 0.6875 & 37.01\% & 0.5907 & 147.63\% \\
                                  & MMOE & 0.6907 & 32.30\% & 0.6857 & 38.34\% & 0.6104 & 103.44\% \\
                                  & PLE & 0.6842 & 36.97\% & 0.6813 & 41.70\% & \underline{0.6375} & 63.35\% \\
                                  & STAR & 0.6874 & 34.63\% & 0.6831 & 40.31\% & 0.6242 & 80.84\% \\
                                  & SAR-Net & 0.6824 & 38.32\% & 0.6763 & 45.72\% & 0.6055 & 112.89\% \\
                                  & DFFM & 0.6973 & 27.88\% & 0.6856 & 38.42\% & 0.6324 & 69.64\% \\  
    \midrule
    \multirow{1}{*}{LLM-based Multi-domain} & \mname & \textbf{0.7523$^\star$} & - & \textbf{0.7569$^\star$} & - & \textbf{0.7246$^\star$} & - \\ 
    \bottomrule
    \end{tabular}
    \begin{tablenotes}
        \item[*] \small \textbf{
            It is worth noting that an AUC increase of 0.001 can be considered a significant improvement in CTR prediction~\cite{inttower, din, 
            autoint,fibinet}.
        }
    \end{tablenotes}
\end{table*}

\paragraph{\textbf{Multi-Domain Models}}

\begin{itemize}[leftmargin=*]
\item \textbf{Shared Bottom.} Shared Bottom~\cite{sharebottom} employs a neural network architecture to extract shared features across various tasks. Additionally, it utilizes a specialized network layer at the top to model unique characteristics specific to each task.

\item \textbf{MMOE.} MMOE \cite{mmoe} is characterized by a shared network of multiple expert submodels and a central gating mechanism that provides implicit connections across diverse tasks with different label spaces.

\item \textbf{PLE.} Different from MMoE, Progressive Layered Extraction \cite{ple} divides experts into task-common and task-specific experts while extending the model from a single-layer network to multiple layers of experts.

\item \textbf{STAR.} STAR \cite{sheng2021one} presents a star-shaped structure centered on a shared network, where the outputs of the shared model are multiplied by the domain-specific outputs.

\item \textbf{SAR-Net.} SAR-Net \cite{shen2021sar} utilizes attention modules to learn users' cross-scenario interests and employs a scenario-specific linear transformation layer, followed by debias expert networks consisting of scenario-specific and shared experts.

\item \textbf{DFFM.} DFFM~\cite{dffm} incorporates domain-related
information into the parameters of the feature interaction and user
behavior modules, allowing for domain-specific learning of these
two aspects.

\end{itemize}

\subsubsection{Implementation Details}
\label{sssec:implementation}

\paragraph{Implementation of \mname}
For the LLM backbone, we employ the Sheared-LLaMA~\cite{shearedllama} architecture, which comprises 1.3 billion parameters and 24 transformer layers. For each domain-specific network (DSN), we employ four ladder layers, each consisting of a small transformer encoder block. 
For the tower network in the DSN, we use three perceptron layers with the dimension of $\{512 \times 256 \times 128 \}$.

\paragraph{Implementation of Baselines.}
For single-domain baselines, the dimension of hidden layers of all MLP classifiers as $\{512 \times 256 \times 128\}$, and other settings can be seen in our open source code.
For the multi-domain baselines in Table~\ref{tab:performance}, we summarize their key configurations as follows:
\begin{itemize}[leftmargin=*, itemsep=0pt, topsep=0pt]
\item Shared Bottom~\cite{sharebottom}: The shared bottom layers have dimensions of $\{1024 \times 1024\}$, whereas the dimensions of the domain-specific layers are $\{1024 \times 512 \times 256\}$.
\item MMOE~\cite{mmoe}: 
The dimensions of tower and expert network are $\{1024 \times 1024\}$ and $\{1024 \times 512\}$, respectively. Note that we empirically configure to deploy $3$ experts for best performance.
\item PLE~\cite{ple}: 
The extraction network of PLE consists $1$ shared expert and $2$ domain-specific experts for each domain. The dimensions of tower and expert network are $\{1024 \times 1024\}$ and $\{1024 \times 512\}$, respectively.
\item STAR~\cite{sheng2021one}: 
The hidden state dimensions of MLPs in auxiliary and star topology networks are $\{1024 \times 512 \times 256\}$.
\item SAR-Net~\cite{shen2021sar}: 
contains $2$ shared and $6$ specific Debias experts, while the hidden state dimension of each expert is $\{1024 \times 1024\}$. The dimension of MLP classifiers is $\{1024 \times 512\}$.

\item DFFM~\cite{dffm}: 
The dimension of the MLP used for Domain Facilitated Feature Interaction (DFFI) is $\{1024 \times 512 \times 256\}$.
\end{itemize}

\paragraph{Zero-Shot Setting}
\label{para:zero-shot-setting}

In the zero-shot experiment, we compare our \mname with six single-domain and four multi-domain models using three training domains (Fashion, Musical Instruments, and Gift Cards), and one test domain (All Beauty). 1) For each single-domain baseline, we train three models with an identical design on the three training domains separately. 
Subsequently, we use these three models to predict the 4$^{th}$ (unknown) domain, resulting in three sets of results. Finally, we choose the best result as the final result of this single-domain baseline. 2) For multi-domain baselines, we train the models on the three training domains.
Although these baselines are implemented with multi-domain classifiers, each for a specific domain, they lack a general network structure specifically designed for out-of-domain prediction.
To mitigate this issue, we use all three classifiers to predict the 4$^{th}$ domain, resulting in three sets of results.
Similar to the approach used for single-domain baselines, we choose the best result as the final result. 3) \mname diverges from the baseline models by utilizing a general network for out-of-domain prediction.
The general network is trained across all three aforementioned domains.
Note that the predictions from three DSNs of \mname are not considered in the out-of-domain scenario.

\paragraph{Optimization and Training}

For \mname, we employ 8 Tesla V100 GPUs with a batch size of 128. To reduce overfitting, we set the dropout~\cite{dropout} rate to 0.3 and utilize L2 regularization~\cite{l2}. The AdamW~\cite{adam} optimizer is used for \mname, and we adopt a Cyclic Learning Rate (CyclicLR) scheme to fluctuate the learning rate between the range of $[1 \times 10^{-6}, 8 \times 10^{-5}]$
To accelerate the training of the LLM backbone, we adopt LoRA~\cite{lora} with a low rank of 8 and an alpha value of 32. 
Regarding the optimizer and hyper-parameter selection of our baselines, we follow the default settings mentioned in their original papers, if applicable.
Otherwise, we employ the Adam~\cite{adam} optimizer with the learning rate of $1 \times 10^{-3}$ and adjust individually for other hyper-parameters. 
We try our best to reproduce their works to obtain the best results, ensuring fair comparisons.

\subsection{Main Results (RQ1)}
\label{ssec:performance-comparison}

In this subsection, we compare the performance of \mname against other single-domain and multi-domain models. The results are summarized in Table~\ref{tab:performance}. From the table, several observations can be obtained:

\begin{itemize}[leftmargin=*]

    \item \textbf{Insights from Single-Domain Models and Existing Multi-domain Models.} The performance of single-domain models, such as DeepFM and xDeepFM, is surpassed by existing multi-domain models in the Fashion and Musical Instruments domains. This demonstrates the effectiveness of single-domain models in data-rich domains. However, single-domain models are significantly inferior to existing multi-domain models in data-sparse domains such as Gift Cards. This suggests that the joint modeling of multiple domains is beneficial for improving the performance of models in sparse domains. However, traditional multi-domain models still suffer from a serious seesaw problem, where data-rich domains (e.g., Fashion and Musical Instruments) still perform significantly better than data-sparse domains (e.g., Gift Cards).

    \item \textbf{Superiority of \mname.} 
    \mname achieved substantial gains in performance across three domains, exhibiting relative improvements of \textbf{24.22\%}, \textbf{35.71\%}, and \textbf{63.35\%} with respect to the AUC metric \footnote{It is worth noting that an AUC increase of 0.001 can be considered a significant improvement in CTR prediction} on Fashion, Musical Instruments, and Gift Cards, respectively. We attribute this to the powerful semantic understanding of the LLM and the powerful characteristic modeling capabilities of DSNs across various domains. Additionally, \mname effectively addresses the seesaw problem caused by data sparsity. In the Gift Cards domain, \mname achieves a significant improvement with a margin of 63.35\%, substantially surpassing the performance gains observed in the other two data-rich domains. We attribute this enhancement to the pre-existing world knowledge embedded in the LLMs. This pre-trained semantic knowledge effectively compensates for the seesaw problem caused by sparse data in the third domain. Furthermore, the observed performance indicates that the inherent world knowledge and semantic understanding capabilities of the LLMs can provide significant cold-start capabilities in domains where data sparsity may hinder model performance. In the subsequent section, this also inspires us to explore the ``zero-shot'' capabilities of our model, assessing its potential utility and effectiveness in previously unseen domains.

\end{itemize}

\subsection{Zero-Shot (Cold Start) Prediction (RQ2)}
\label{ssec:zero-shot-prediction}

\begin{figure}[htbp]
    \centering
    \includegraphics[width=0.9\linewidth]{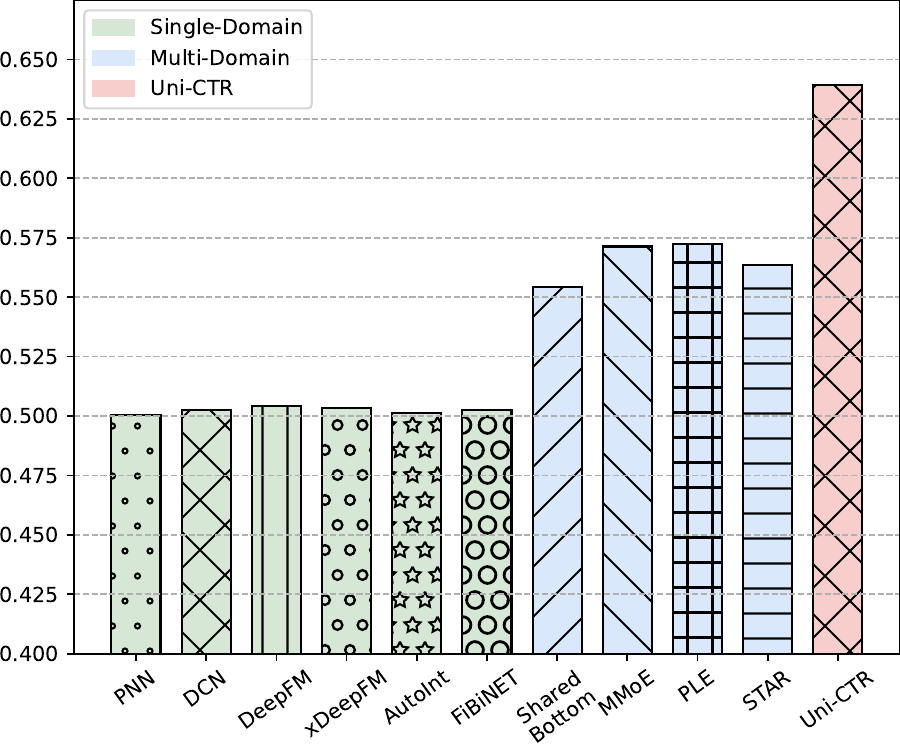}
    \caption{Comparative performance of zero-shot prediction on traditional models and \mname on the unseen domain (All Beauty).}
    \label{fig:zero-shot-comparison}
\end{figure}

In real-world industrial recommendation systems, it is common for new business domains to emerge, often accompanied by new, unknown items. This typical situation is referred to as the cold-start problem in recommendation systems. In such cases, we lack sufficient training data to train the model. Consequently, the capability of existing models to cope with cold-start problems is of utmost importance. Therefore, this section investigates the model's performance under a zero-shot setting. The experiment is set up in Section~\ref{para:zero-shot-setting}. As illustrated in Fig.~\ref{fig:zero-shot-comparison}, the single-domain models demonstrate the limited capacity to generalize to new domains, as evidenced by their low AUC values of around 0.51 or below. These models are limited to understanding and predicting only within the data distribution of the domains in which they are trained, resulting in poor performance in zero-shot prediction.

For multi-domain models, we observe a marked enhancement in zero-shot prediction capabilities. We hypothesize that this improvement stems from the model's ability to leverage common information across multiple domains during training, which allows it to generalize to previously unseen domains.

Our \mname model outperforms all the baseline models, with a notable improvement exceeding \textbf{6\%} points compared to traditional multi-domain models. This superior performance can be attributed to two main factors: (i) incorporating world knowledge embedded within the LLM aids in bolstering the multi-domain model's performance under a cold-start setting, and (ii) the General Network in \mname effectively distills common knowledge across multiple domains, ensuring robust generalization capabilities when confronted with new domains.

\subsection{LLM Scale Up (RQ3)}
\label{ssec:scale-up}

\begin{figure*}[t]
    \centering
    \includegraphics[width=0.9\linewidth]{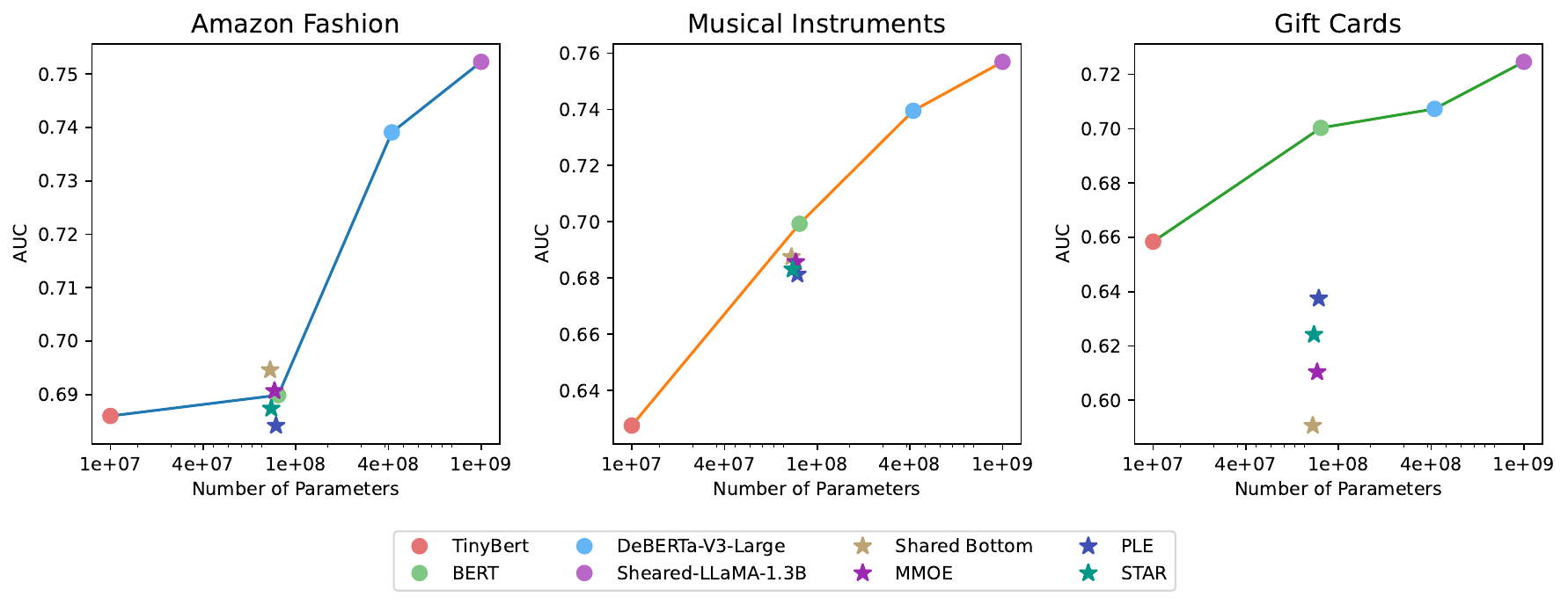}
    \caption{Performance comparison of different language model backbones.}
    \label{fig:scaleup}
\end{figure*}

The scaling law~\cite{scalinglaw} is a pivotal principle in training LLMs, describing how various aspects of a model and its training process change with scale, such as model size, dataset size, and computational resources. Specifically, the more parameters a model has and the more extensive training data it is fed, the better performance it exhibits. However, inference latency is also paramount in recommendation systems in the industry. For the CTR prediction task, the inference latency per request is typically constrained to within $10$ milliseconds. Therefore, in this subsection, we examine the impact of parameter counts in the LLM backbone on model performance. We employ four language models as backbones: TinyBERT~\cite{tinybert}, BERT~\cite{bert}, DeBERTa-V3-Large~\cite{DeBERTaV3}, and Sheared-LLama~\cite{shearedllama} with respective parameter counts of 14M, 110M, 340M, and 1.3 billion. 

The experimental results are summarized in Fig.~\ref{fig:scaleup}, from which we obtain some observations: 
1) With the increasing size of LLMs, a notable enhancement in performance is observed, which indicates that the scaling laws are equally applicable to \mname. Furthermore, the LLM backbone of \mname primarily captures the commonalities across domains,  which provides a potential avenue for enhancing the effectiveness of traditional multi-domain CTR models in the future. 
2) \mname, based on a BERT Backbone with 110M parameters, has already surpassed traditional multi-domain models. This illustrates the powerful semantic comprehension and world knowledge embedded within language models, leading to significant performance gains. Thus, our model provides a viable approach for application within the industry recommender systems. 


\subsection{Scalability (RQ4)}
\label{ssec:scalability}

\begin{table}[htbp]
    \caption{Performance comparison of different models when scaled to a new domain.}
    \label{tab:scalability}

    \resizebox{\linewidth}{!}{%
        \begin{tabular}{ccccc}
        \hline
        \multirow{2}{*}{Category}      & \multirow{2}{*}{Models} & \multirow{2}{*}{Scalability} & \multicolumn{2}{c}{Digital Music}      \\ \cline{4-5} 
                                       &                         &                              & AUC                              & RelaImpr \\ \hline
        \multirow{6}{*}{Single-domain} & PNN                     & $\times$                     & 0.5904                           & 26.11\%  \\
                                       & DCN                     & $\times$                     & 0.5919                           & 24.05\%  \\
                                       & DeepFM                  & $\times$                     & 0.5917                           & 24.32\%  \\
                                       & xDeepFM                 & $\times$                     & 0.5957                           & 19.12\%  \\
                                       & AutoInt                 & $\times$                     & 0.5913                           & 24.86\%  \\
                                       & FiBiNET                 & $\times$                     & 0.5832                           & 37.02\%  \\ \hline
        Multi-domain                   & Shared Bottom           & $\checkmark$                 & 0.5975                           & 16.92\%  \\
                                       & STAR                    & $\checkmark$                 & 0.6038                           & 9.830\%  \\
                                       & MMOE,PLE                & $\times$                     & -                                & -        \\ \hline
        LLM-based Multi-domain         & \mname                  & $\checkmark$                 & \textbf{0.6140}                  & -        \\ \hline
        \end{tabular}
    }
\end{table}

To evaluate the scalability of the model, we freeze the weights of \mname\, which has already been trained on three domains: Fashion, Musical Instruments, and Gift Cards. We then add an additional DSN for fine-tuning the Uni-CTR on a new domain, Digital Music. During training, only the parameters of the newly added DSN are updated. For single-domain models, we retrain it on the Digital Music domain. As for a scalable model like STAR, we also add a new network structure to scale it, and the training method is consistent with Uni-CTR.
The results are presented in Table~\ref{tab:scalability}.

We can observe that by merely training an additional Domain-Specific Network (DSN), the performance of Uni-CTR is improved by 9.830\% compared to the multi-domain STAR model, and at least 19.12\% compared to the fully retrained single-domain baseline model. This evidence confirms that the \mname has successfully identified and assimilated the commonalities across different domains during its prior training. Such an understanding of these commonalities facilitates the adaptation of \mname to new domains, utilizing this prior knowledge as a foundational base. Therefore, even with the original parameters of \mname frozen, the newly added DSN demonstrates impressive performance. This efficiency not only conserves significant computational resources but also substantially reduces the time required for model training.

\subsection{Visualization (RQ5)}
\label{ssec:visualization}

In this subsection, we use t-distributed Stochastic Neighbor Embedding (t-SNE)~\cite{tsne} to visualize representations of LLMs as well as representations of DSNs in latent space to explore their role in modeling multi-domain commonalities and characteristics.
 
    


\begin{figure*}[htbp]
    \centering
    \subfloat[$\boldsymbol{h}_0$]{%
        \includegraphics[width=0.3\textwidth]{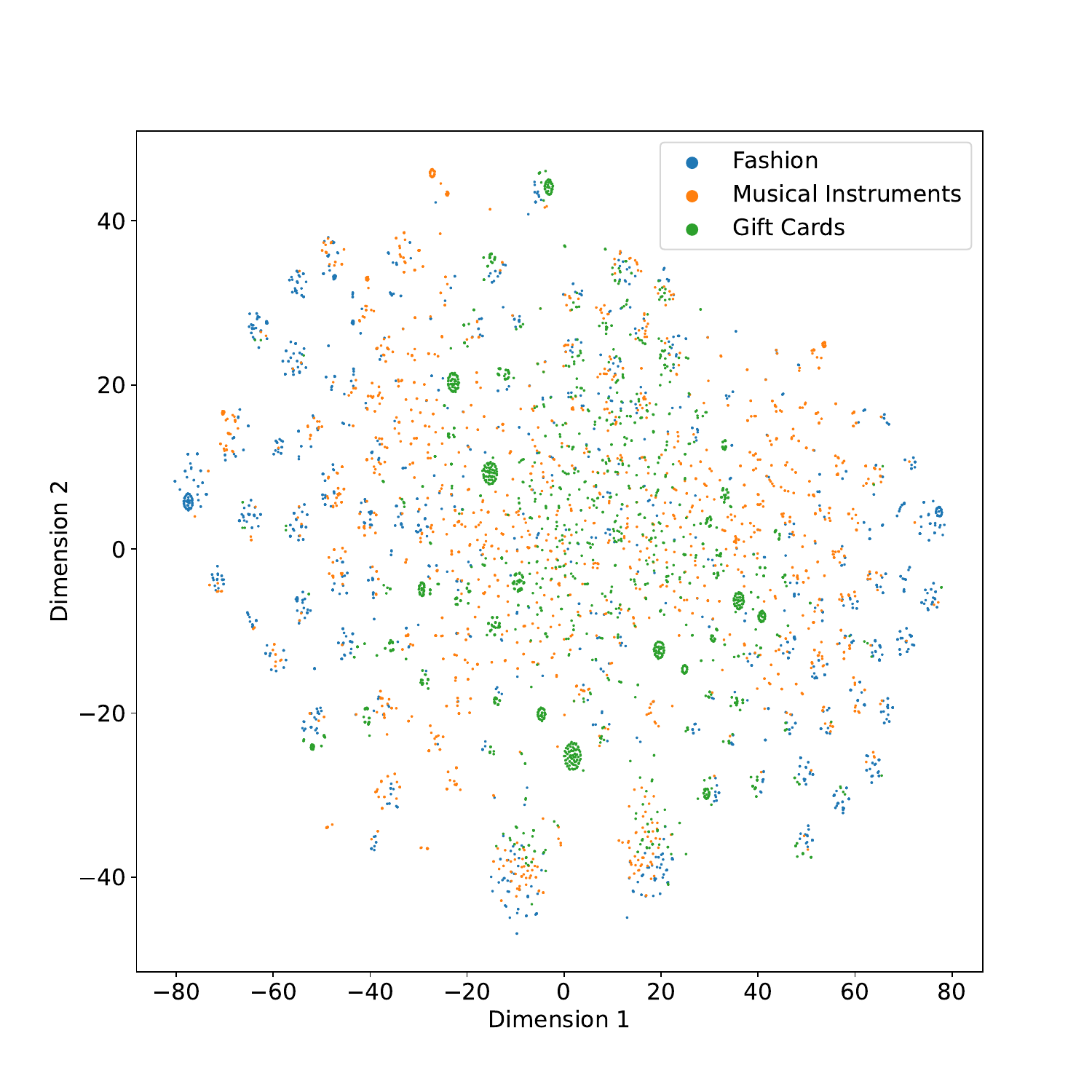}
        \label{subfig:llm-untrained0}
    }\hfill 
    \subfloat[$\boldsymbol{h}_2$]{%
        \includegraphics[width=0.3\textwidth]{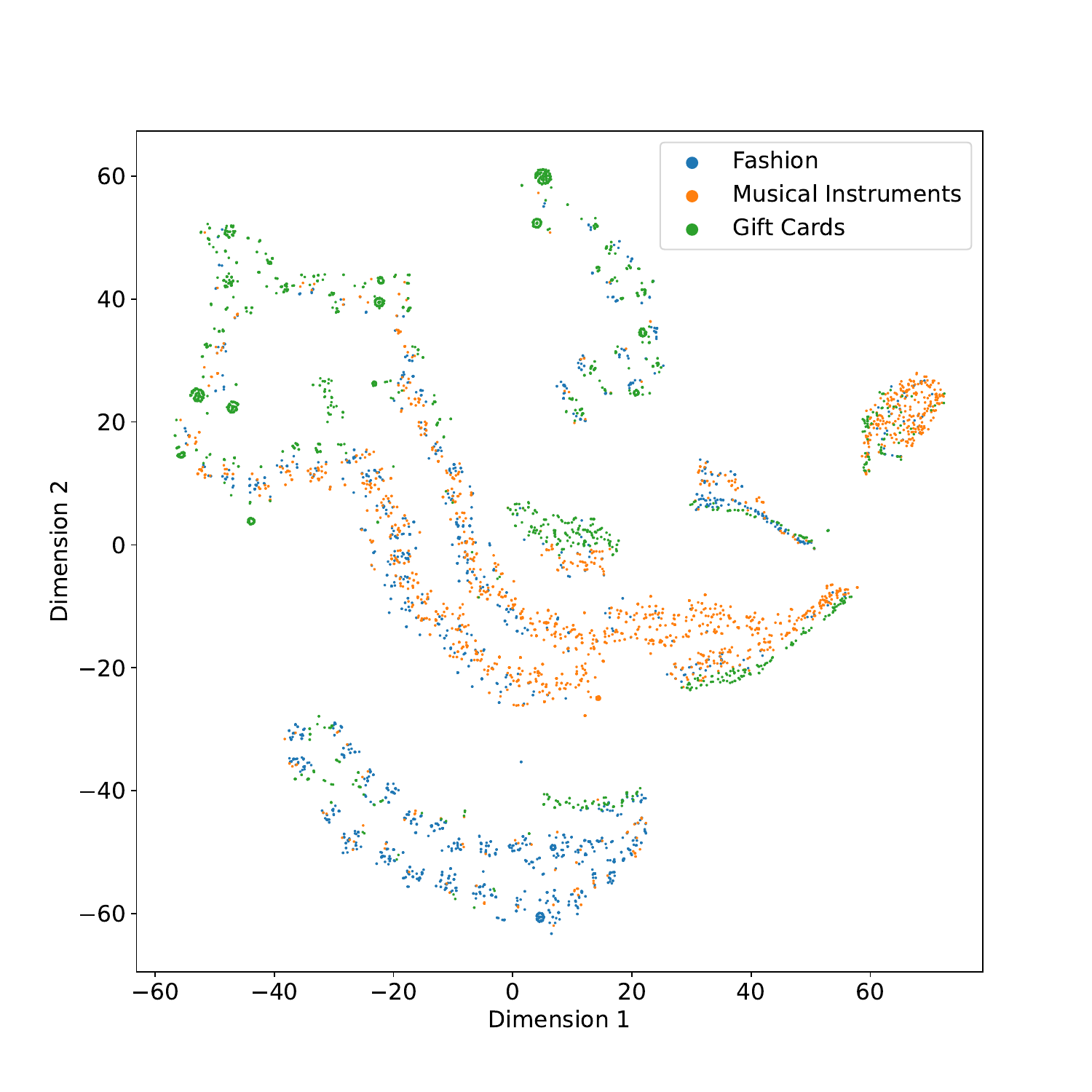}
        \label{subfig:llm-untrained1}
    }\hfill 
    \subfloat[$\boldsymbol{h}_4$]{%
        \includegraphics[width=0.3\textwidth]{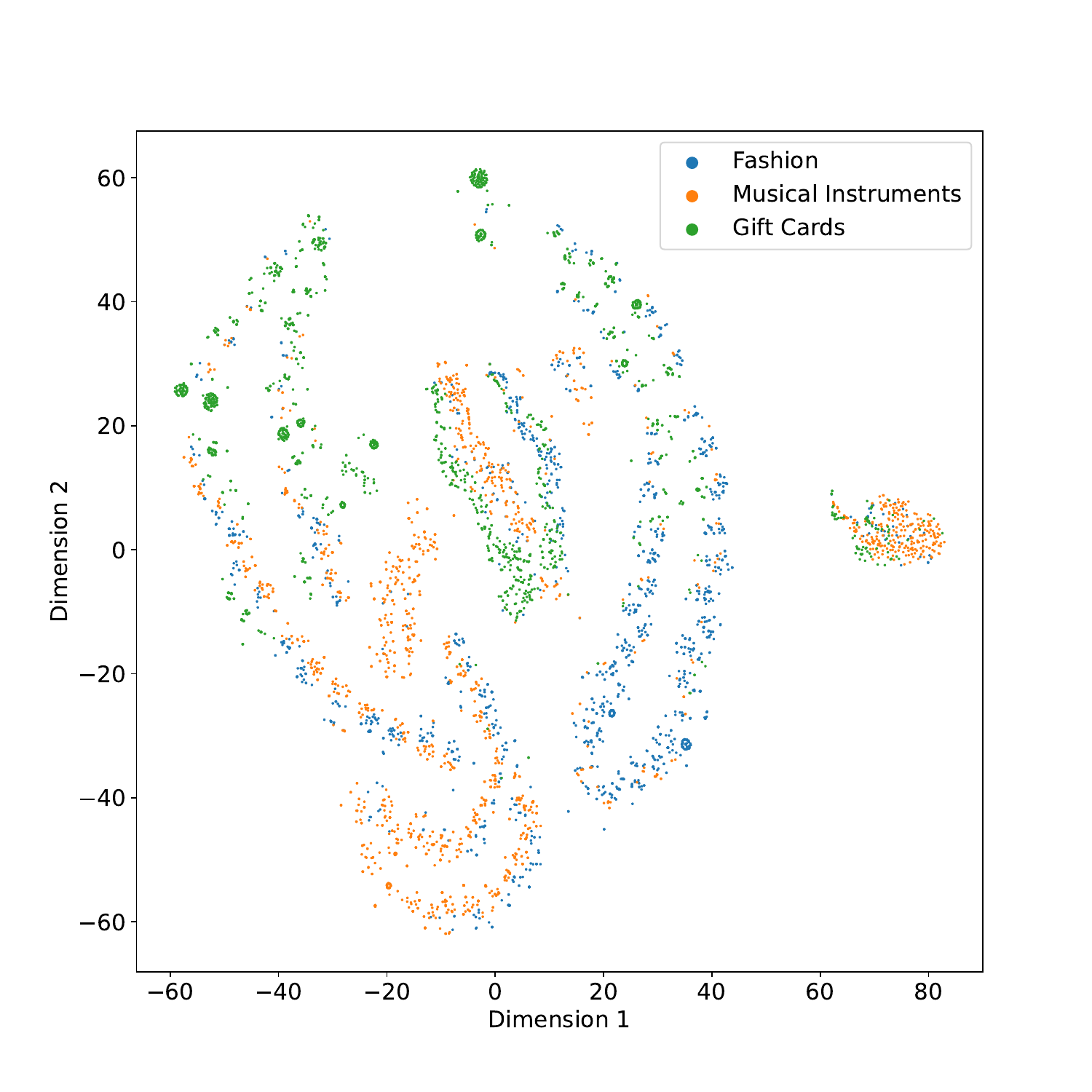}
        \label{subfig:llm-untrained2}
    }
    \caption{Visualization of the representations of different layers of the LLM.}
    \label{fig:visualization-llm}
\end{figure*}

\begin{figure*}[htbp]
    \centering
    \subfloat[Untrained DSN]{%
        \includegraphics[width=0.4\textwidth]{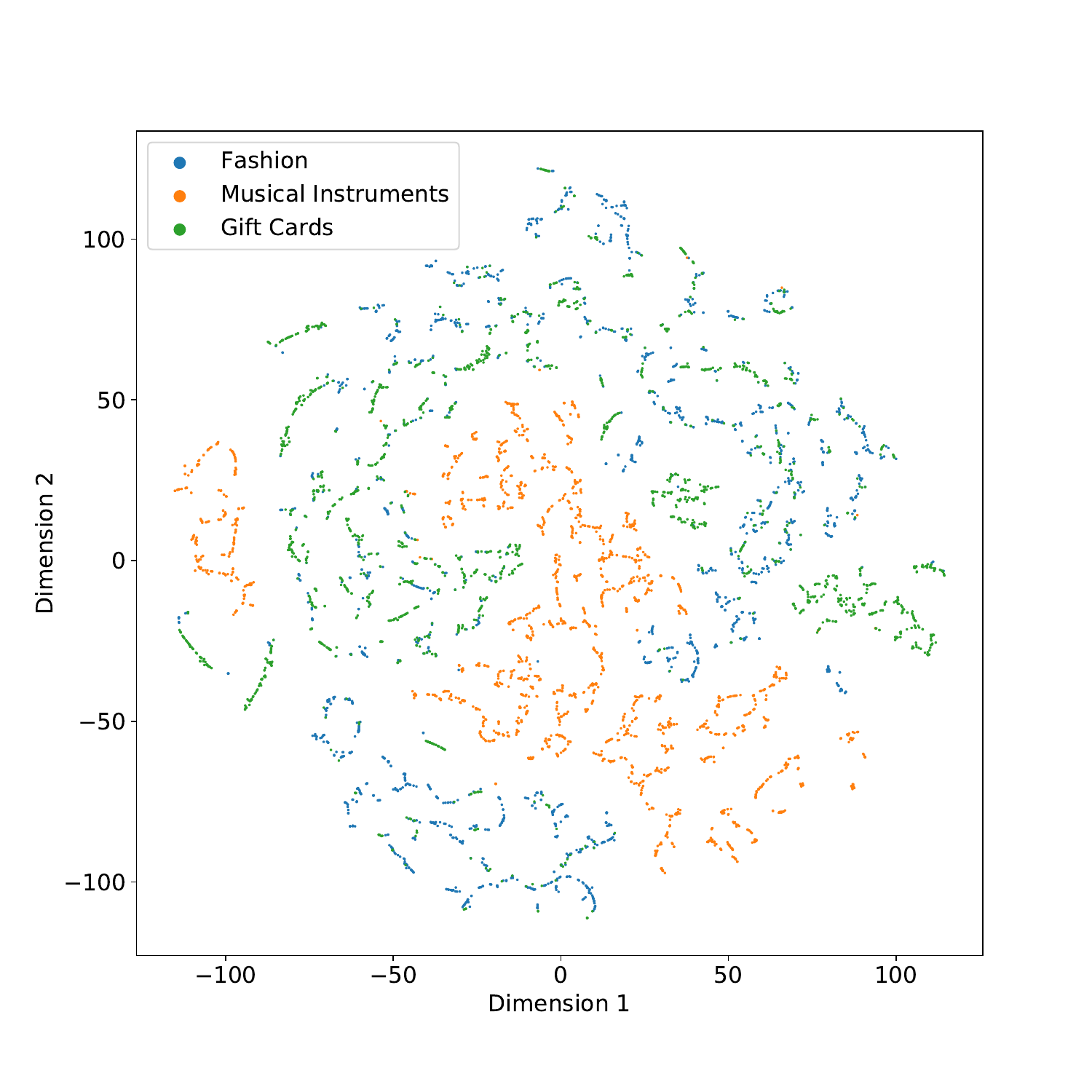}
        \label{subfig:dsn-untrained}
    }
    \hspace{1cm}
    \subfloat[Trained DSN]{%
        \includegraphics[width=0.4\textwidth]{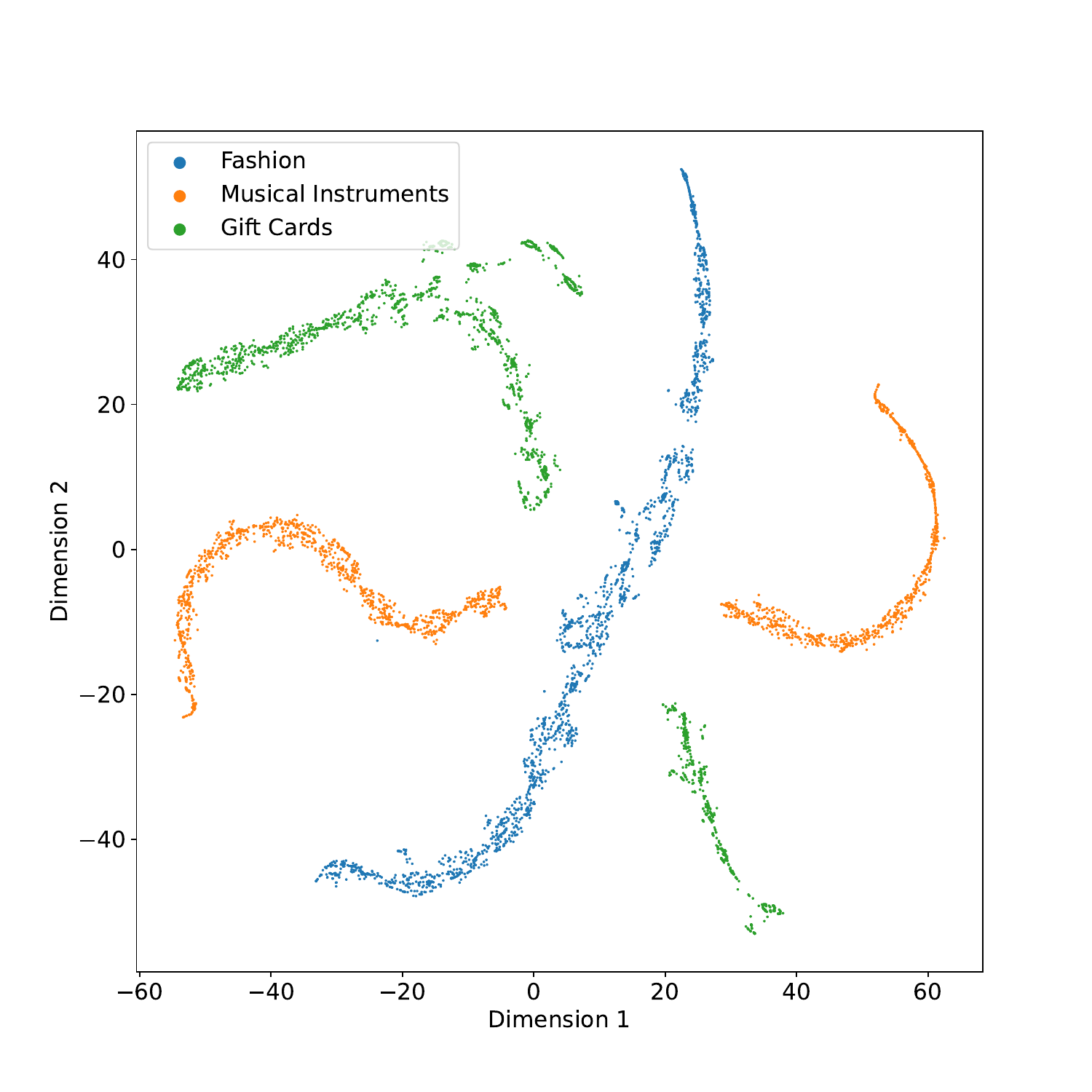}
        \label{subfig:dsn-trained}
    }
    \caption{Visualization of the representations of DSNs.}
    \label{fig:visualization-dsn}
\end{figure*}

\subsubsection{Visualization of LLM Representations} 

As depicted in Fig.~\ref{fig:visualization-llm}, we visualize the representations from the $0^{th}$, $2^{nd}$, and $4^{th}$ layers of the LLM, denoted as $h_0$, $h_2$, and $h_4$, respectively.
These representations are illustrated as colored dots, with each color corresponding to a different training domain.

From Fig.~\ref{fig:visualization-llm}, we can observe two significant results. 
1) The representations at different transformer layers of the LLM are distinctly distributed in the latent space, indicating that different layers of the LLM capture semantic representations at various levels. Such finding is in good agreement with the results of previous studies~\cite{jawahar-etal-2019-bert,rogers2021primer}. Therefore, it is necessary to incorporate ladder networks at different LLM layers to model semantic information at multiple levels, a point further substantiated in Section~\ref{ssec:ablation}. 
2) As illustrated in Fig.~\ref{subfig:llm-untrained0}, representations from various domains at the lower layers of LLM are completely aggregated. While at higher layers, as shown in Fig.~\ref{subfig:llm-untrained1} and Fig.~\ref{subfig:llm-untrained2}, the representations of different domains gradually start to separate. This demonstrates that the LLM is capable of not only modeling the commonalities among multiple domains but also capturing the coarse-grained characteristics of each domain, which helps the DSN network further extract information about the characteristics of the domains.

\subsubsection{Visualization of DSN Representations} 
In order to validate the role of DSNs, we extract the representations from the penultimate layer of the tower network within DSN for each domain, comparing their states before and after training, as shown in Fig.~\ref{fig:visualization-dsn}. 
From Fig.~\ref{subfig:dsn-untrained}, we can observe that the distributions of the representations of the three domains are mixed in the latent space before the DSNs are trained. This indicates that untrained DSNs are incapable of effectively distinguishing the representations of different domains, i.e., they cannot accurately model the characteristics of each domain. After training, as shown in Fig.~\ref{subfig:dsn-trained}, the representations of different domains in the latent space are well-separated. This shift demonstrates that DSNs gain the ability to distinguish the representations of various domains and accurately capture the characteristics of each domain through training.

\subsection{Ablation Study (RQ6)}
\label{ssec:ablation}

To clarify the role of different components in the \mname model, 
we conduct ablation studies and the results are summarized in Table~\ref{tab:ablation-prompt} and Table~\ref{tab:ablation-ladder-llm}.

\subsubsection{Impact of Prompt}

\begin{table}[htbp]
    \caption{Ablation Study Results about the input text of \mname(BackBone DeBERTa-V3-Large).}
    \label{tab:ablation-prompt}
    \centering
    \resizebox{\linewidth}{!}{%
        \begin{tabular}{@{}cccc@{}}
            \toprule
            Input Text      & Fashion        & Musical Instruments & Gift Cards \\ \midrule
            Full Prompt          & 0.7047         & 0.7008            & 0.6825     \\
            Only Feature ID and Feature Name    & 0.6960         & 0.6960            & 0.6749     \\
             Only Feature ID         & 0.6951         & 0.6835            & 0.6605     \\ \bottomrule
        \end{tabular}
    }
\end{table}

To assess the importance of prompt semantics in LLMs, we explore the impact of different prompts on model performance, as detailed in Table~\ref{tab:ablation-prompt}. The `Full Prompt' format, the original design for \mname discussed in Section~\ref{ssec:prompt}, utilizes complete semantic information. The `Only Feature ID and Feature Name' format combines feature names and feature ID into a single string, removing other semantic information. The `Only Feature ID' format further simplifies this by retaining only feature ID values. We compared the results after training the same 5 epochs using \mname with DeBERTa-V3-Large as the backbone LLM. Our findings indicate a marked decline in performance with the reduction of semantic information in the prompts.

The analysis of the results reveals the significance of prompt design in LLMs. In all three domains (Fashion, Musical Instruments, and Gift Cards), the fully semantic `Full Prompt' format consistently outperformed the other two. Specifically, the performance dropped by 1.23\% and 1.36\% in the `Ony Feature ID and Feature Name' and `Only Feature ID' formats, respectively, for Fashion domain. 
Moreover, the most pronounced decline was observed in the Gift Cards domain compared to the other two domains. This significant drop can be attributed to the relatively smaller size of the dataset in this domain. The relatively small data size necessitates a higher reliance on rich semantic information to compensate for the lack of extensive data points, making the impact of prompt design more pronounced.

These results emphasize the critical role of semantic information in enhancing LLMs' understanding and prediction accuracy. The performance of the model decreases as the semantic information decreases, suggesting that LLM relies heavily on semantic information to make accurate predictions.  Thus, while simplifying the input format can improve computational efficiency, it clearly affects the performance of the model. Considering these findings, optimizing prompt design is a crucial aspect when deploying LLMs for MDCTR prediction.

\subsubsection{Impact of Ladder Network and LLM}

\begin{table}[htbp]
    \caption{Ablation Study Results of \mname (BackBone DeBERTa-V3-Large).}
    \label{tab:ablation-ladder-llm}
    \centering
    \begin{tabular}{@{}cccc@{}}
        \toprule
        Model      & Fashion        & Musical Instruments & Gift Cards \\ \midrule
        \mname     & 0.7391         & 0.7395            & 0.7073     \\
        w/o ladder & 0.7084         & 0.6975            & 0.6723     \\
        w/o LLM    & 0.6954         & 0.6923            & 0.6100     \\
        MMOE(340M) & 0.7038         & 0.7005            & 0.6712     \\
        STAR(340M) & 0.7107         & 0.7016            & 0.6775     \\ \bottomrule
    \end{tabular}
\end{table}

In order to clarify the role of domain-specific networks as well as the LLM Backone in the \mname model, we conduct the following ablation study, and the results are summarized in Table~\ref{tab:ablation-ladder-llm}.
1) We remove the ladder network, restricting the architecture to only utilize the LLM and the three Tower layers for both training and prediction. We observe a considerable degradation in performance, indicative of the essential role that the ladder network plays in leveraging multi-level semantic information from LLM. This suggests that the extraction of characteristic information pertinent to distinct domains benefits from the intricate semantics provided by the ladder network, and relying solely on the LLM is insufficient. 
2) We remove the LLM, i.e., we replace the LLM with a DNN with the same number of layers, and the inputs are changed to the traditional IDs. we can observe that when the LLM backbone is lost, not only does the model show a huge loss of performance on all domains, but also the drop is significant on data-sparse domains (Gifts Cards). This shows that LLM plays the most crucial role in improving the performance on data-sparse domains. 
3) To verify whether it is the increase in the number of parameters that brings about the performance improvement, we increase the number of parameters of the models of MMOE and PLE to the same size as that of Uni-CTR (340M). The results show that even with the same number of parameters, the performance of the traditional recommendation model is still much weaker than that of Uni-CTR, but better than that of Uni-CTR without LLM or ladder layer, which indicates that both the ladder layer and LLM backbone contribute to the performance of Uni-CTR.

\section{Industrial Experiments}
\label{sec:industry}

\begin{table}[t]
    \caption{The overall performance of models trained on the industrial dataset.}
    \label{tab:industry}
    \resizebox{\linewidth}{!}{%
        \begin{tabular}{@{}cccccc@{}}
        \toprule
        \multirow{2}{*}{Category} & \multirow{2}{*}{Models} & \multicolumn{2}{c}{Domain 0} & \multicolumn{2}{c}{Domain 1} \\ \cmidrule(l){3-6} 
        &  & AUC & RelaImpr & AUC & RelaImpr \\ \midrule
        \multirow{6}{*}{Single-domain} & PNN & 0.6735 & 37.58\% & 0.6199 & 56.88\% \\
        & DCN & 0.6722 & 38.62\% & 0.6243 & 51.33\% \\
        & DeepFM & 0.6743 & 36.95\% & 0.6223 & 53.80\% \\
        & xDeepFM & 0.6738 & 37.34\% & 0.6226 & 53.43\% \\
        & AutoInt & 0.6788 & 33.50\% & 0.6214 & 54.94\% \\
        & FiBiNET & 0.6780 & 34.10\% & 0.6146 & 64.14\% \\ 
        \midrule
        \multirow{3}{*}{Multi-domain} & MMoE & 0.7045 & 16.72\% & 0.6640 & 14.70\% \\
        & PLE & 0.7019 & 18.23\% & 0.6706 & 10.26\% \\
        & STAR & 0.7000 & 19.35\% & 0.6638 & 14.84\% \\
        \midrule
        LLM-based Multi-domain & \mname & \textbf{0.7387} & - & \textbf{0.6881} & - \\ 
        \bottomrule
        \end{tabular}%
    }
\end{table}

This section outlines the practical application and empirical evaluation of \mname in a real-world industrial setting. 
In preparation for the experiment, 
we gather and sample one month of user behavior data from a large-scale industrial recommender system.
This platform generates millions of user logs daily, providing a substantial and diverse collection of data on user interactions and preferences.

\subsection{Model Performance of Industry Platform}

In this real-world application, users are divided into two domains based on business requirements, referred to as \textit{Domain 0} and \textit{Domain 1}.
The single-domain models are individually trained on each domain, while the multi-domain models and our proposed \mname are trained conjointly on both domains.

The results are summarized in Table~\ref{tab:industry}. From the table, we observe the following phenomena: 
1) In this industrial scenario, even multi-domain models outperform single-domain models by a large margin. We speculate that this is because modeling the commonality of multiple domains is particularly important in large-scale industrial datasets. However, in smaller datasets such as Amazon, the impact of multi-domain commonalities on various domains is not obvious. 
2) Trained on real-world industrial datasets, \mname outperforms SOTA multi-domain models, achieving an impressive relative improvement in AUC of over \textbf{10.26\%}. This significant performance gain can be attributed to the adoption of text-based input, which not only enhances the model's flexibility but also enriches its semantic understanding. Such rich semantic understanding proves to be a crucial factor in the superior performance of \mname.

\subsection{Model Inference Acceleration}

In industrial recommender systems, the online model serving latency is subject to a strict constraint, typically set at around 1 to 2 milliseconds for a single instance. As a result, ensuring high service efficiency holds paramount importance for CTR models. However, for applications using LLMs, latency~\cite{li2023ctrl} is an intractable problem because of the complex attention mechanism and the excessive depth of the transformer layers. 

For inference acceleration, we export the trained model (with backbone DeBERTa-V3-Large) to .onnx\footnote{\url{https://onnx.ai/}} format, which enables us to perform model inference utilizing a static graph paradigm. We then quantize it using FP16 precision with the assistance of the TensorRT tool\footnote{\url{https://github.com/NVIDIA/TensorRT}}. With a batch size of 32 and a sequence length of 256, the \mname inference latency on a single Tesla V100 GPU is 80ms. The average per-sample latency is around 2ms, and the loss in AUC is below 0.01, which is still significantly better than existing traditional multi-domain recommendation models. This latency is perfectly acceptable for \mname to be used in the rank stage for industrial recommender systems.
\section{Conclusion}
\label{sec:conclusion}

In this paper, we propose \mname, a unified framework for multi-domain CTR prediction. 
It comprises an LLM backbone plugged with multiple domain-specific networks and a general network. 
The introduced LLM backbone learns common features across various domains from the designed prompts with its powerful semantic understanding. 
After, the domain-specific networks receive layer-wise representations from the transformer layers of the LLM to capture characteristics inherent to each specific domain. 
Simultaneously, the general network learns common patterns across all the domains to enable zero-shot prediction for unseen domains. 
In the extensive experiments conducted on both public and industrial datasets, the \mname model outperforms existing single-domain and multi-domain models, effectively mitigating the ``seesaw phenomenon'' and improving generalization to new domains. 
Future research will continue to investigate the enhancement in input modalities for multi-domain CTR prediction.










\section*{Acknowledgment}

This research was partially supported by Huawei (Huawei Innovation Research Program), Research Impact Fund (No.R1015-23), APRC - CityU New Research Initiatives (No.9610565, Start-up Grant for New Faculty of City University of Hong Kong), CityU - HKIDS Early Career Research Grant (No.9360163), Hong Kong ITC Innovation and Technology Fund Midstream Research Programme for Universities Project (No.ITS/034/22MS), Hong Kong Environmental and Conservation Fund (No. 88/2022), and SIRG - CityU Strategic Interdisciplinary Research Grant (No.7020046, No.7020074).

\bibliographystyle{IEEEtran}
\bibliography{sample-base}

\end{document}